%% file: main.tex
\documentclass[twocolumn,times]{aastex631}

\usepackage{graphicx}
\usepackage{epstopdf}
\usepackage{subfigure}
\usepackage{color}
\usepackage{soul}
\usepackage{tabularx}
\usepackage{booktabs}
\usepackage{amsmath}
\usepackage{appendix}
\let\tablenum\relax
\usepackage{siunitx}
\usepackage{enumitem}
\usepackage{CJK}
\usepackage[normalem]{ulem}
\usepackage{cancel}

\newcommand{\um}{$\mu$m}
\newcommand{\UMGAl}{COS-DR1-209435}
\newcommand{\UMGBl}{COS-DR3-195616}
\newcommand{\UMGA}{209435}
\newcommand{\UMGB}{195616}
\newcommand{\logM}{$\log (M_{\star}/M_{\odot})$}

\newcommand{\zphot}{$z_\textrm{phot}$}
\newcommand{\zspec}{$z_\textrm{spec}$}

\newcommand{\OII}{\hbox{{\rm [O}\kern 0.1em{\sc ii}{\rm ]$\lambda\lambda3726,3729$}}}
\newcommand{\OIIIlater}{\hbox{{\rm [O}\kern 0.1em{\sc iii}{\rm ]$\lambda5007$}}}
\newcommand{\OIIIfront}{\hbox{{\rm [O}\kern 0.1em{\sc iii}{\rm ]$\lambda4959$}}}

\newcommand{\Hbeta}{$\rm{H}\beta$}
\newcommand{\Halpha}{$\rm{H}\alpha$}
\newcommand{\NIIlater}{\hbox{{\rm [N}\kern 0.1em{\sc ii}{\rm ]$\lambda6584$}}}

\shortauthors{Chang et al.}
\graphicspath{{./}{figures/}}

\begin{document}

\title{MAGAZ3NE: Far-IR and Radio Insights into the Nature and Properties of Ultramassive Galaxies at $z\gtrsim3$}

\correspondingauthor{Wenjun Chang}
\email{wenjun.chang@email.ucr.edu}

\author[0000-0003-2144-2943]{Wenjun Chang}
\affiliation{Department of Physics and Astronomy, University of California, Riverside, 900 University Avenue, Riverside, CA 92521, USA}

\author[0000-0002-6572-7089]{Gillian Wilson}
\affiliation{Department of Physics, University of California, Merced, 5200 Lake Road, Merced, CA 95343, USA}

\author[0000-0001-6003-0541]{Ben Forrest}
\affiliation{Department of Physics and Astronomy, University of California, Davis, One Shields Avenue, Davis, CA 95616, USA}

\author[0000-0002-2446-8770]{Ian McConachie}
\affiliation{Department of Astronomy, University of Wisconsin-Madison, 475 N. Charter St., Madison, WI 53706 USA}

\author{Tracy Webb}
\affiliation{Department of Physics, McGill Space Institute, McGill University, 3600 rue University, Montr\'{e}al, Qu\'{e}bec H3A 2T8, Canada}

\author[0000-0003-1832-4137]{Allison Noble}
\affiliation{School of Earth and Space Exploration, Arizona State University, Tempe, AZ 85287, USA}

\author[0000-0002-9330-9108]{Adam Muzzin}
\affiliation{Department of Physics and Astronomy, York University, 4700, Keele Street, Toronto, ON MJ3 1P3, Canada}

\author[0000-0003-1371-6019]{M. C. Cooper}
\affiliation{Center for Cosmology, Department of Physics and Astronomy, University of California, Irvine, Irvine, CA, USA}

\author[0000-0001-9002-3502]{Danilo Marchesini}
\affiliation{Department of Physics \& Astronomy, Tufts University, MA 02155, USA}

\author[0000-0003-4693-6157]{Gabriela Canalizo}
\affiliation{Department of Physics and Astronomy, University of California, Riverside, 900 University Avenue, Riverside, CA 92521, USA}

\author[0000-0003-4569-2285]{A. J. Battisti}
\affil{International Centre for Radio Astronomy Research, University of Western Australia, 35 Stirling Hwy, Crawley, WA 6009, Australia}
\affil{Research School of Astronomy and Astrophysics, Australian National University, Cotter Road, Weston Creek, ACT 2611, Australia}

\author[0000-0002-9466-2763]{Aurelien Henry}
\affiliation{Department of Physics, University of California, Merced, 5200 Lake Road, Merced, CA 95343,  USA}  

\author{Percy Gomez}
\affiliation{W.M. Keck Observatory, 65-1120 Mamalahoa Hwy., Kamuela, HI 96743, USA}

\author[0000-0001-8169-7249]{Stephanie M. Urbano Stawinski}
\affiliation{Center for Cosmology, Department of Physics and Astronomy, University of California, Irvine, Irvine, CA, USA}

\author[0000-0002-6505-9981]{M.E. Wisz}
\affiliation{Department of Physics, University of California, Merced, 5200 Lake Road, Merced, CA 95343, USA} 
 


\begin{abstract}


Deep and wide-field near-infrared (NIR) surveys have recently discovered and confirmed ultramassive galaxies (UMGs; \logM\ $>11$) spectroscopically at high redshift. However, most are characterized using only ultraviolet (UV)-to-NIR photometry, offering limited insight into obscured star formation and active galactic nucleus (AGN) activity.
In this work, we add ten far-infrared (FIR)-to-radio passbands to the existing UV-to-NIR catalogs for two spectroscopically confirmed UMGs from the MAGAZ3NE survey, \UMGBl\ (\mbox{$z_{\rm spec} = 3.255$}) and \UMGAl\ (\mbox{$z_{\rm spec} = 2.481$}).
Utilizing the full UV-to-radio photometry, we revise our earlier UV-NIR-based interpretation of the nature of these galaxies. While both were previously identified as quiescent, our analysis reveals that \UMGB\ is an unobscured galaxy undergoing quenching, and \UMGA\ is a heavily obscured, actively star-forming UMG.   
We find that \UMGB\ has already depleted most of its molecular gas and is expected to experience minimal future stellar mass growth. In contrast, \UMGA\ contains a substantial molecular gas reservoir and has a prolonged depletion timescale.  It is anticipated to increase 0.34 dex in stellar mass, reaching a stellar mass of \mbox{\logM\ = 11.72} over 
the next 0.72~Gyr. We present multi-pronged evidence for AGN activity in both UMGs. Our findings support a scenario where AGN feedback in \UMGB\ may have contributed to gas depletion during quenching, while \UMGA\ continues to form stars despite hosting an obscured AGN, suggesting feedback has not yet suppressed star formation.  
Our work shows the importance of FIR-to-radio observations for accurately inferring the nature and properties of galaxies at $z\gtrsim3$.
 
\end{abstract}

\keywords{Far infrared astronomy (529); Galaxy Evolution (594); High-redshift Galaxies (734); Molecular Gas (1073); Star Formation (1569); AGN Host Galaxies (2017); Galaxy Quenching (2040)}

\section{Introduction} \label{sec:intro}

Ultramassive galaxies (UMGs; \mbox{\logM\ $>$ 11}) lie at the most extreme tail of the galaxy stellar mass function and offer critical insights into the physical processes driving early galaxy formation and evolution.
Observing these systems at high redshift allows us to capture key formation phases, such as episodes of intense star formation followed by rapid quenching \citep{Belli2014, Schreiber2018a_J&H, Forrest2020a, Forrest2020b}. These phases are no longer accessible in the local Universe, as galaxies have already assembled most of their stellar mass and have long since shut down their star formation, making it difficult to directly trace their formative processes \citep{Muzzin2013b, Marsan2017}.
Over the last decade, numerous studies using deep near-infrared (NIR) multi-wavelength surveys have led to the photometric identification of a growing number of massive galaxies at higher redshifts (e.g., \citealt{Stefanon2015, Marsan2022, Carnall2023, Labbe2023}).
However, securing spectroscopic confirmation of these relatively rare and faint massive galaxies has proven challenging.
A multi-year Keck/MOSFIRE campaign, the ``Massive Ancient Galaxies At $z >$ 3 NEar-infrared'' (MAGAZ3NE) survey, has been successful in spectroscopically confirming a sample of UMGs, photometrically-selected to lie at \mbox{$3\lesssim$~\zphot~$<4$}. MAGAZ3NE spectroscopic and photometric samples have been used to characterize UMG properties using multi-passband ultraviolet (UV)-to-NIR catalogs both directly by the MAGAZ3NE survey team and through its extended collaborations (e.g., \citealt{Forrest2020a, Saracco2020, Forrest2022, Forrest2023, Forrest2024a, Forrest2024b, Marsan2022, McConachie2022, McConachie2025, Stawinski2024}; McConachie et al.\ in prep; Urbano Stawinski et al.\ in prep; Chang et al.\ in prep). 
  
The MAGAZ3NE survey has fundamentally reshaped our understanding of the field of massive galaxy formation, but also revealed the limitations of UV-NIR-only observations. Many MAGAZ3NE UMGs, for instance, exhibit extremely red spectral energy distributions (SEDs) \citep{Forrest2024b}, but without far-infrared (FIR) or radio constraints, it remains unclear whether those red colors indicate truly quiescent systems or heavily dust-obscured star-forming galaxies, especially in cases where emission lines are weak or attenuated. 
This distinction is crucial. Dusty star-forming galaxies (DSFGs) are known to dominate the high-mass end of the galaxy population (\mbox{\logM\ $>10.3$}) at $z>2$ \citep{Martis2016}, and recent studies suggest that their numbers at $z>3$ have been significantly underestimated by optical-NIR surveys alone \citep{Wang2019, Barrufet2023}. As a result, relying solely on UV-NIR observations increases the risk of misclassifying DSFGs as quenched systems.
Furthermore, although \cite{Forrest2020b} identified signatures of active galactic nucleus (AGN) activity in some UMGs 
using \OIIIlater/\Hbeta\ diagnostics, UV-NIR observations alone are insufficient to robustly quantify the full AGN contribution, particularly from heavily obscured sources. To conclusively distinguish quiescent galaxies from dust-obscured starbursts and to assess the role of AGN, it is essential to incorporate longer wavelength observations.


In this work, we combine public FIR-to-radio observations in the COSMOS/UltraVISTA field with the existing UV-to-NIR photometric catalogs to investigate the nature and properties of spectroscopically-confirmed UMGs from the MAGAZ3NE survey in the early Universe. This comprehensive UV-to-radio analysis enables a more complete characterization of their star formation, gas content, and AGN activity, and provides new insights into their evolutionary pathways.
Our work is organized as follows: Section\ \ref{sec:survey} summarizes the MAGAZ3NE survey, while Section\ \ref{sec:catalogs} describes the sample selection and multi-wavelength photometric catalog creation. Section\ \ref{sec:analysis} describes the SED fitting, and estimation of stellar mass, star formation rate, molecular gas, depletion timescale and stellar mass evolution. 
A discussion of the results is presented in Section\ \ref{sec: Discussion}, and a summary of the conclusions in Section\ \ref{sec: conc}.
Throughout this work we assume a \citet{Chabrier2003} initial mass function (IMF) and a $\Lambda$CDM cosmology with $H_0=70$~km~s$^{-1}$~Mpc$^{-1}$, $\Omega_M=0.3$, and $\Omega_\Lambda=0.7$. We utilize the AB magnitude system \citep{Oke1983}.


\section{MAGAZ3NE Survey}
\label{sec:survey}


MAGAZ3NE survey is a spectroscopic follow-up survey of candidate UMGs, selected from UltraVISTA DR1 \citep{Muzzin2013a}, UltraVISTA DR3 \citep{Marsan2022}, and XMM-VIDEO \citep{Jarvis2013}.
Candidates were selected to have a photometric redshift of \zphot~$\geq~3$, a photometric stellar mass of \mbox{\logM~$>11.0$}, 
and a well-sampled photometric UV-to-NIR SED. The first phase of the survey targeted candidates with a photometric redshift of \mbox{$3\lesssim$~\zphot~$<4$} and a photometric stellar mass of \mbox{$11.0<$\logM$<11.7$}.
This target population consisted almost entirely of either blue star-forming or post-starburst galaxies, and resulted in 16 robust spectroscopic redshifts. Excellent agreement was found between the spectroscopic and photometric redshifts and, therefore, between the spectroscopically- and photometrically-inferred stellar masses \citep{Forrest2020b}.

The second phase of the MAGAZ3NE survey also targeted candidates at a photometric redshift of \mbox{$3\lesssim$~\zphot~$<4$}, but focused on candidates with the highest 
photometrically-estimated stellar mass, those with photometric stellar mass of \logM\ $>11.7$, a population that we termed Super-ultramassive galaxies (S-UMGs).
This S-UMG population consisted of candidates with very red SEDs, implying significant dust attenuation, old stellar ages, and/or AGN. This phase of the survey resulted in 12 robust spectroscopic redshifts (\citealt{Forrest2024b}). However, in contrast to the good agreement between spectroscopic and photometric redshifts in the first phase of the survey, in this second phase of the survey, there was a significant discrepancy between the spectroscopically- and photometrically-estimated redshifts, primarily from overestimation of the photometric redshifts.
In fact, no member of the S-UMG population turned out to have a \mbox{$\lvert$~\zspec-\zphot~$\rvert <0.5$}. \cite{Forrest2024b} discussed the universal difficulty that photometric redshift programs appear to have in fitting galaxies with very red SEDs.  
 
The poor agreement between spectroscopic and photometric redshifts for MAGAZ3NE candidates with red SED also led to discrepancies in stellar mass estimates.
Although all of the candidates were confirmed to be massive, none of the candidates were confirmed to be a {\it bona fide} S-UMG at \mbox{$3\lesssim$~\zphot~$<4$}. 
This had important implications for galaxy evolution, suggesting that the number densities of high-mass galaxies determined from large photometric surveys at $z\gtrsim3$ are overestimated (see \citet{Forrest2024b} for further discussion). 
 
\begin{figure}
\centering
\includegraphics[width = 0.48\textwidth]{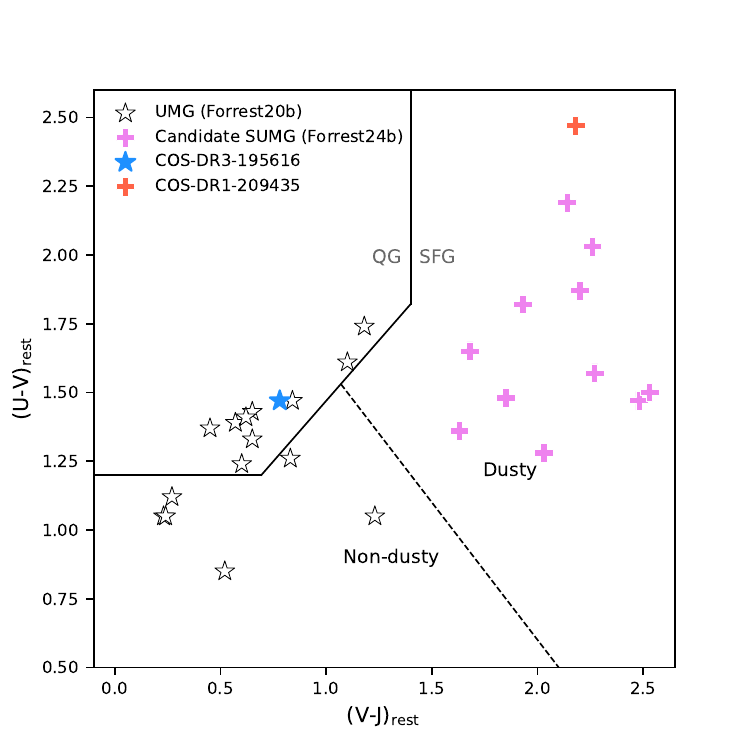}
\caption{ 
Rest-frame \textit{UVJ} color
of the 16 spectroscopically-confirmed UMG candidates presented in \cite{Forrest2020b} and 12 spectroscopically-confirmed S-UMGs candidates presented in \cite{Forrest2024b}. 
The RF colors were estimated at the spectroscopic redshift using FAST$++$ and UV-to-8\um\ photometry only.  
The dashed line denotes the division between  ``dusty'' and ``non-dusty'' star-forming regions \citep{Schreiber2018b_UVJ_FAST}.  
\UMGBl\ lies in the quiescent ``wedge" 
and \UMGAl\ lies in the upper right dusty star-forming region. 
\label{fig: UVJ}}
\end{figure}

Figure\ \ref{fig: UVJ} shows the rest-frame \textit{U$-$V} and \textit{V$-$J} (RF \textit{UVJ}) colors of the 28 spectroscopically-confirmed MAGAZ3NE UMGs (white stars and magenta crosses).\footnote{
Figure\ \ref{fig: UVJ} shows the (RF) \textit{UVJ} colors of the 28 MAGAZ3NE candidate UMGs with confirmed spectroscopic redshifts. Nine of the 28 S-UMG candidates turned out to have either a \mbox{$z_{\rm spec} < 3.0$} or a stellar mass of \mbox{\logM$ < 11.0$} (see Table 1 in \citealt{Forrest2020b} and Table 2 in \citealt{Forrest2024b}).  
For conciseness, however, in this work we refer to all members of the sample as UMGs.}   
The RF \textit{U$-$V} and \textit{V$-$J} colors were derived by \citet{Forrest2020b, Forrest2024b} using UV-to-8\um\ photometry (with $K$-band corrected for emission line flux) and fitted by FAST$++$ \citep{Kriek2009, Schreiber2018b_UVJ_FAST} at the spectroscopic redshift of each galaxy. Full details can be found in \citet{Forrest2020b} and  \citet{Forrest2024b}.  
In Section\ \ref{sec:catalogs}, we match public ALMA and other long-wavelength observations to the sample of 15 spectroscopically-confirmed MAGAZ3NE UMGs in the COSMOS UltraVISTA field. We find matches for two MAGAZ3NE UMGs (blue star and red cross in Figure\ \ref{fig: UVJ}), and add FIR-to-radio passbands to the existing 
UV-to-NIR catalogs to create two new UV-to-radio catalogs.


\section{Multiwavelength Catalog Creation in the COSMOS/UltraVISTA Field} 
\label{sec:catalogs} 

\subsection{Parent catalogs: UltraVISTA DR1 and DR3}
\label{subsec: DR1_DR3}
 
As the parent photometric catalogs for the MAGAZ3NE survey, the UltraVISTA DR1 and DR3 catalogs are constructed from the UltraVISTA survey \citep{McCracken2012}, which has four deep near-infrared $Y$-, $J$-, $H$-, and $K_{s}$-band imaging over 1.62 deg$^{\rm 2}$ in the COSMOS field. \cite{Muzzin2013a} combined the UltraVISTA Data Release One (DR1) with additional photometry from 0.15-24 \um, yielding a total of 30 bandpasses 
with 90\% completeness at $K_{s}$ = 23.4 mag.

Further deep imaging in the NIR from Data Release Three (DR3) imaged the ``ultra-deep stripes'' with a total area of approximately 0.84 deg$^{2}$. The DR3 photometric catalog was constructed following the same procedure as outlined in \cite{Muzzin2013a} and consists of a total of 49 passbands. The NIR depths are deeper than DR1 by $\sim$1.2 mag, reaching $K_{s}$ = 25.2 AB mag (5$\sigma$, 2.1\arcsec diameter aperture; \citealt{Marsan2022}), and are $\sim$1 mag deeper in the IRAC 3.6 and 4.5 \um\ bandpasses \citep{Ashby2018}.

\subsection{ALMA A$^{3}$COSMOS Catalog and Sample Selection}
\label{subsec: ALMA_A3}

Of the 28 confirmed UMGs and S-UMGs in the MAGAZ3NE sample, 15 are located within the COSMOS/UltraVISTA field. To explore the available ALMA data in this field, we adopt the public A$^{3}$COSMOS catalog of \citet{Liu2019, Liu2019b}, created by automated mining of the ALMA Archive in the COSMOS field. 
Within a search radius of 0.5\arcsec, we cross-match the 15 MAGAZ3NE UMGs with the A$^3$COSMOS (Dataset 20200310\footnote{A$^3$COSMOS Dataset 20200310 \url{https://sites.google.com/view/a3cosmos/data/dataset_v20200310?authuser=0}.}) prior-extracted photometric catalog (see \citealt{Liu2019} for details). 
This search identifies two matches in Band 7 (872 \um) exceeding the 3$\sigma$ threshold (Flux$_{\rm peak}/$RMS): \UMGBl\ (from program 2016.1.00463.S, PI: Y. Matsuda; angular resolution 0.69\arcsec) and \UMGAl\ (from program 2015.1.00137.S, PI: N. Scoville; angular resolution 0.78\arcsec), which we refer to hereafter, for brevity, as \UMGB\ and \UMGA, respectively. 
As can be seen from Figure~\ref{fig: UVJ}, 
\UMGB\ (blue star) falls in the quiescent ``wedge'' \citep{2011Whitaker} of the \textit{UVJ} diagram, suggesting that it is either a passive galaxy or a galaxy undergoing quenching, and \UMGA\ (red cross) falls in the upper right ``red'' corner, a region known to contain dusty and/or AGN-dominated galaxies.

\begin{figure*}
\centering
\includegraphics[width = 0.98\textwidth]{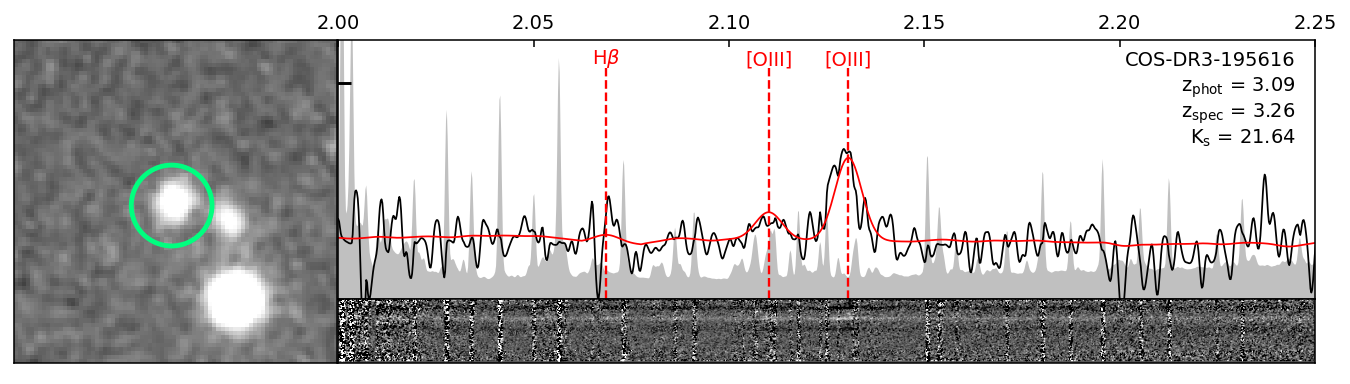}
\includegraphics[width = 0.98\textwidth]{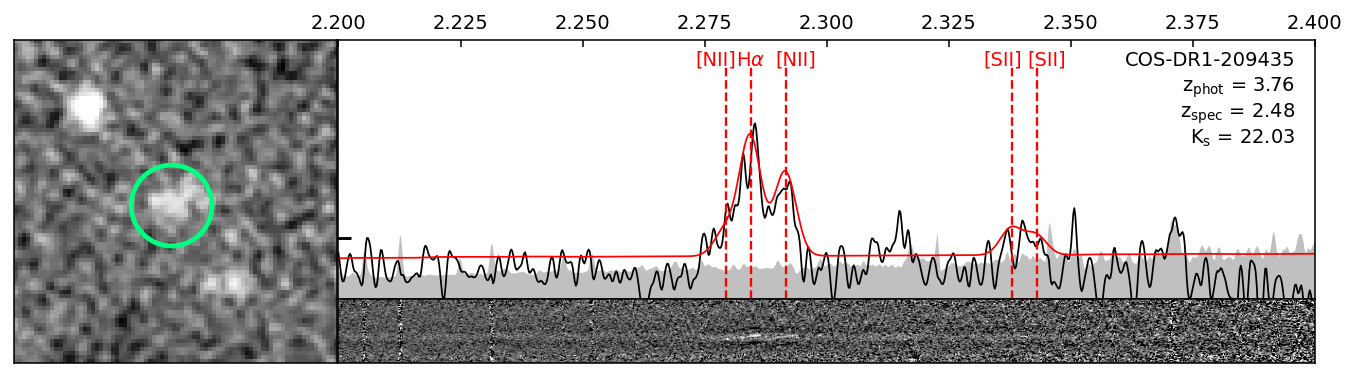}
\caption{K$_{s}$-band images (left) and MOSFIRE 1D and 2D $K$-band spectra (right) for \UMGBl\ (upper) and \UMGAl\ (lower). The black solid lines show the spectra smoothed over 5 pixels and weighted by the inverse variance. The gray shading shows the magnitude of the error spectrum. The solid red lines show the best-fit models from FAST$++$
described in Section\ \ref{subsec: redshift}. The green circles are drawn with a 2\arcsec diameter.
Red vertical dashed lines indicate the location of prominent emission lines. The y-axis shows flux density in units of \mbox{$10^{-18}$ erg s$^{-1}$ cm$^{-2}$ \AA$^{-1}$}, plotted from zero with a dash indicating one unit.
\label{fig: Ks_img_spectra}}
\end{figure*}

\subsection{Spectroscopic Observations}
\label{subsec: redshift}

Figure~\ref{fig: Ks_img_spectra} shows the $K_{s}$-band images and Keck/MOSFIRE 1D and 2D $K$-band spectra for \UMGB\ and \UMGA.
We fit the UV-to-NIR photometry and MOSFIRE spectroscopy using the combination of \texttt{slinefit}\footnote{https://github.com/cschreib/slinefit} and FAST$++$ to determine the best-fit spectroscopic redshift ($z_{\mathrm{spec}}$) and emission line fluxes, correcting the photometry for emission-line contributions and masking those lines in the spectra. (see \S4 in \citealt{Forrest2020b} for details).
The solid red line in Figure~\ref{fig: Ks_img_spectra} shows the best-fit model from \citet{Forrest2024b}, obtained by combining the multi-Gaussian fit to the emission lines with the stellar continuum model from FAST$++$. The FAST$++$ fit was performed with emission lines masked and the photometry corrected for line contamination, ensuring a clean continuum estimate.
UMG \UMGB\ was first presented in \citet{Forrest2020b}.
As with other spectroscopically-confirmed members of the MAGAZ3NE sample found in the quiescent wedge of the \textit{UVJ} diagram, the photometric (\mbox{$z_{\rm phot}=3.09^{+0.09}_{-0.08}$}) and spectroscopic (\mbox{$z_{\rm spec}= 3.255^{+0.001}_{-0.001}$}) redshifts of \UMGB\ are generally consistent within 3$\sigma$ uncertainties. 
In contrast, as with other spectroscopically-confirmed ``red'' members of the MAGAZ3NE sample presented in \citet{Forrest2024b}, the spectroscopic redshift of \UMGA\ (\mbox{$z_{\rm spec}=2.481^{+0.001}_{-0.001}$}) significantly disagrees with its photometric redshift (\mbox{$z_{\rm phot}=3.76^{+0.34}_{-0.34}$}), even considering the substantial uncertainty in $z{\rm phot}$. Throughout this paper, ``red'' refers to a UMG with a very red RF \textit{V$-$J} color (\textit{V$-$J} $>2$), implying significant dust attenuation. 
The presence of an AGN in \UMGB\ was inferred by \citet{Forrest2020b} from its position on the Mass-Excitation diagram, which compares the line ratio \OIIIlater/\Hbeta\ with the stellar mass \citep{Juneau2011}. We will discuss the AGN contribution further in Section \ref{subsec: AGN_frac}.

\subsection{FIR/Radio Observations}
\label{subsec: FIR_radio_Jin18}

The very large beam sizes of FIR/(sub)mm detectors cause source confusion (blending), making it challenging to accurately measure galaxy fluxes at those wavelengths. \citet{Jin2018} applied the ``super-deblending” technique developed by \citet{Liu2018} to create a 15-passband COSMOS catalog (hereafter as Jin18) consisting of \textit{Spitzer}/IRAC 3.6, 4.5, 5.8, and 8.0 \um\ (SPLASH; PI: P. Capak), MIPS 24 \um\ (COSMOS-\textit{Spitzer}; PI: D. Danders; \citealt{LeFloch2009}), \textit{Herschel}/PACS 100 and 160 \um\ (PEP, PI: D. Lutz; \citealt{Lutz2011} and CANDELS-\textit{Herschel}, PI: M.Dickinson), \textit{Herschel}/SPIRE 250, 350, and 500 \um\ (HerMES; PI: S. Oliver), SCUBA2 850 $\mu$m (S2CLS; \citealt{Cowie2017, Geach2017}), AzTEC 1.1 mm \citep{Aretxaga2011}, MAMBO 1.2 mm \citep{Bertoldi2007}, and VLA 3 GHz and 1.4 GHz \citep{Smolvcic2017, Schinnerer2010}. 
We find detections in the Jin18 catalog within 0.5\arcsec\ of both \UMGB\ and \UMGA\ in the UltraVISTA DR1\&DR3 catalogs.

In creating our own combined multi-passband catalog for each of the two UMGs, we have a choice of utilizing the IRAC and MIPS 24 \um\ observations from either the UltraVISTA DR1\&DR3 or Jin18 catalogs. 
To assess consistency between the two datasets, we cross-match galaxies from the UltraVISTA and Jin18 catalogs and compare the extracted IRAC and MIPS 24 \um\ fluxes for all matched sources. 
In the case of the four IRAC passbands, there is little difference in the photometry between the catalogs, and so we elected to utilize the IRAC photometry from the UltraVISTA catalog to maintain consistency with the UV-to-NIR photometric analysis carried out in \cite{Forrest2020b}. However, in the case of MIPS 24 \um, the fluxes in the Jin18 catalog exceeded those from the DR1 and DR3 catalogs by a median factor of 2.16 and 7.22, respectively. 
This discrepancy in the 24 \um\ photometry between Jin18 and the UltraVISTA DR1 catalogs was also noted by \cite{Jin2018} and attributed to a combination of heavy source blending in the MIPS 24 \um\ images and different calibration methodologies.
Therefore, in creating our multi-passband catalogs, we have determined to utilize the 24 \um\ photometry from the Jin18 catalog, as it employs deblending techniques to recover fluxes in the heavily confused MIPS 24 \um\ images.

The photometric catalogs that we construct and utilize in this analysis contain ten additional FIR-to-radio passbands (shown in Table~\ref{tab:passband}), which are not included in the optical-IR analyses presented in \citet{Forrest2020b} and \citet{Forrest2024b}. Beam FWHM values for the FIR passbands are from \citet{Jin2018}, which represent the full widths at half maximum of the circular Gaussian approximation to the point spread function of each image, except for ALMA, which refers to the angular resolution of the respective programs described in Section~\ref{subsec: ALMA_A3}. Both \UMGB\ and \UMGA\ are observed in all ten of these additional FIR-to-radio passbands.
In Section\ \ref{sec:analysis}, we explore what new information these more comprehensive catalogs reveal.

\begin{deluxetable}{ccc}
\setlength{\tabcolsep}{10pt}
\tablecaption{FIR-to-radio passbands
utilized in this work. 
\label{tab:passband}}
\tablehead{
\colhead{Instrument} & \colhead{Wavelength} & \colhead{Beam FWHM (arcsec)}
}
\startdata
\textit{Spitzer}/MIPS & 24 $\mu$m & 5.7 \\
\textit{Herschel}/PACS & 100 $\mu$m & 7.2\\
\textit{Herschel}/PACS & 160 $\mu$m & 12.0\\
\textit{Herschel}/SPIRE & 200 $\mu$m & 18.2\\
\textit{Herschel}/SPIRE & 350 $\mu$m & 24.9\\
\textit{Herschel}/SPIRE & 500 $\mu$m & 36.3\\
JCMT/SCUBA2 & 850 $\mu$m & 11.0\\
ALMA/band7 & 872 $\mu$m & 0.7\\
VLA/3GHz & 10 cm & 0.8 \\
VLA/1.4GHz & 21.4 cm& 2.5\\
\enddata
\end{deluxetable}

\section{Analysis}
\label{sec:analysis}



\subsection{SED Fitting}
\label{ssec:analysis}

In the following sections, we study how three factors (spectroscopic versus photometric redshift, choice of SED fitting code, and full UV-to-radio versus only UV-to-NIR information)
affect estimates of the key properties of the UMGs (stellar mass, star formation rate, dust extinction, AGN fraction, and age). We compare results from two SED fitting codes: FAST$++$ and \texttt{CIGALE}, 
and present the results for \UMGB\ in Table~\ref{tab:all_DR3} and \UMGA\ in Table~\ref{tab:all_DR1}.
Firstly, we utilize the results from \citet{Forrest2020a, Forrest2024b}, which employed FAST$++$\footnote{\url{https://github.com/cschreib/fastpp}} \citep{Kriek2009, Schreiber2018b_UVJ_FAST} and the UV-to-NIR photometry to estimate the properties of each UMG at the photometric redshift (see Section\ \ref{subsec:FAST} and column 2), and at the spectroscopic redshift (see Section\ \ref{subsec:FAST++} and column 3). 
Next, in this work, we employ \texttt{CIGALE} \citep{Boquien2019} and the UV-to-NIR photometry to estimate the properties of each UMG at the spectroscopic redshift (see Section\ \ref{subsec: CIGALE} and column 4).
Finally, we employ \texttt{CIGALE} and the full UV-to-radio photometry to estimate the properties of each UMG at the spectroscopic redshift (see Section\ \ref{subsec:CIGALEradio} and column 5).
The results from these different approaches are discussed and analyzed in Section\ \ref{ssec:SMSFR}.

\subsubsection{UV-to-NIR SED fitting with FAST$++$ at z$_{phot}$}
\label{subsec:FAST}


We use the UV-to-NIR photometry and the photometric redshift from the DR1 \citep{Muzzin2013a} and DR3 \citep{Marsan2022} UltraVISTA photometric catalogs (detailed in Section\ \ref{subsec: DR1_DR3}). 
The stellar population properties were estimated with FAST$++$ in \citet{Forrest2020b}, assuming the stellar population model from \citep{BruzualCharlot03} with a \citet{Chabrier2003} initial mass function (IMF), a double-exponential SFH, and the starburst dust law from \citet{Calzetti2000}. We show the results for \UMGB\ and \UMGA\ in the second columns of Tables~\ref{tab:all_DR3} and~\ref{tab:all_DR1}.


\input{Table_all_stellar_prop_DR3}
\input{Table_all_stellar_prop_DR1}

\subsubsection{UV-to-NIR SED fitting with FAST$++$ at z$_{spec}$}
\label{subsec:FAST++}

We next employ FAST$++$ \citep{Kriek2009, Schreiber2018b_UVJ_FAST} and the UV-to-NIR photometry to estimate the properties of each UMG at the spectroscopic redshift (further details may be found in \citealt{Forrest2020b}). 
The templates used in the fit are the same as those described in Section~\ref{subsec:FAST}. The resulting estimates are shown in the third column of Tables~\ref{tab:all_DR3} and~\ref{tab:all_DR1}.


\begin{figure*}
\centering
\gridline{\fig{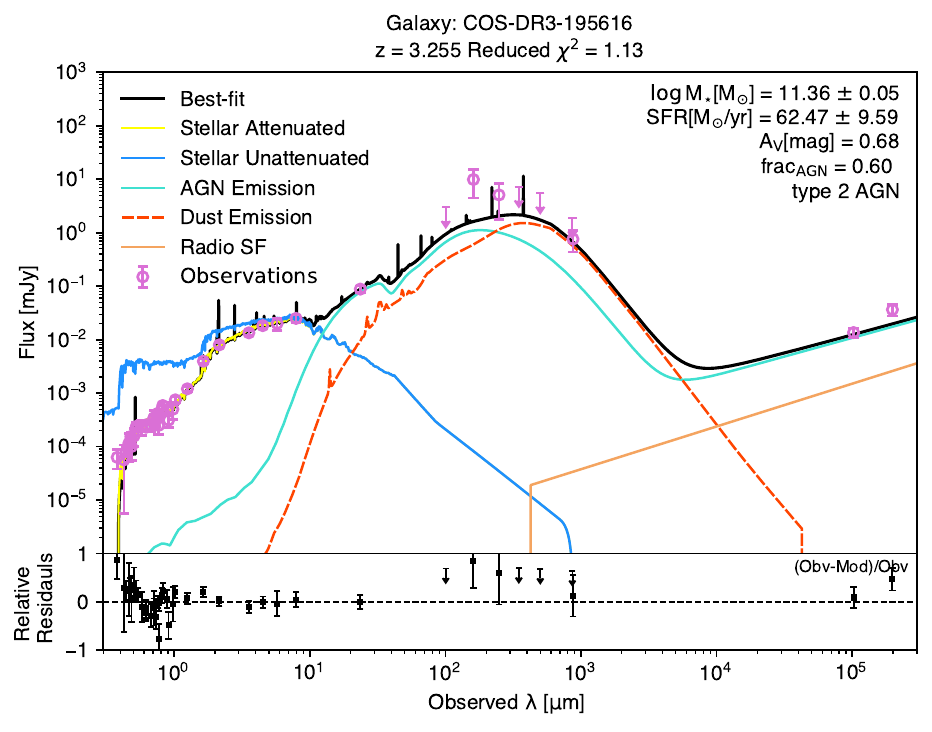}{0.49\textwidth}{(a)}
\fig{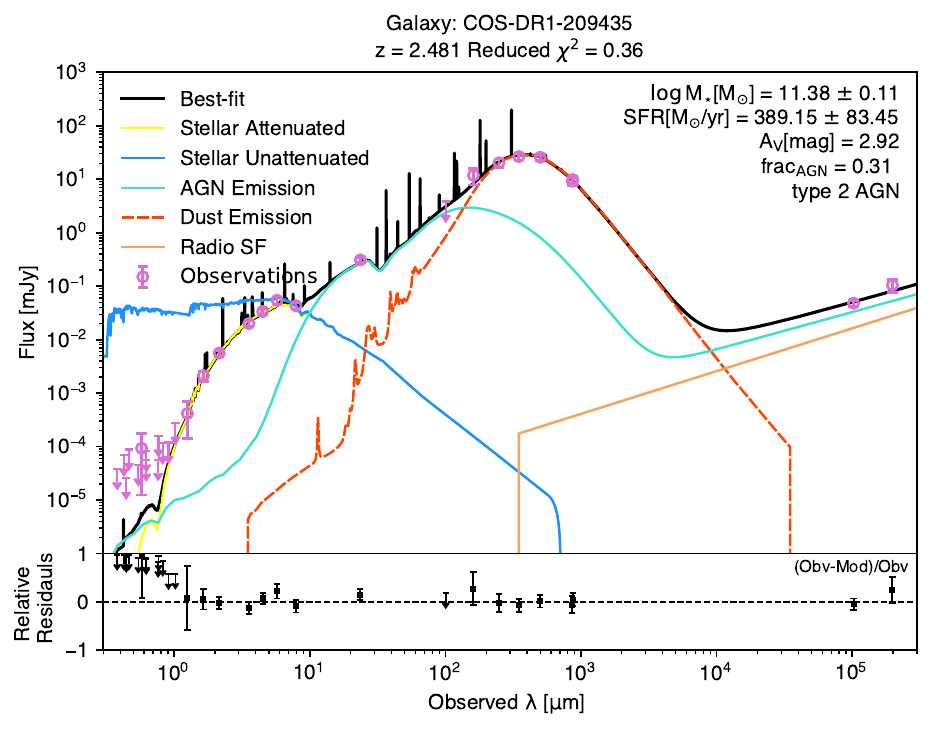}{0.49\textwidth}{(b)} }
\caption{
Best fit SED (black) and residuals for \UMGBl\ (left) and \UMGAl\ (right), estimated by \texttt{CIGALE} from the UV-to-radio catalogs. Also shown is the unattenuated stellar (blue), dust (red), AGN (cyan), and radio synchrotron non-thermal (orange) emission. Observations are shown by purple circles, with downward arrows indicating upper limits.
Estimates of physical properties are shown in the last column of Tables~\ref{tab:all_DR3} and~\ref{tab:all_DR1}. There appears to be a significant AGN contribution in the case of both UMGs (see~Section\ \ref{subsec: AGN_frac}).
\label{fig:CIGALE_two}}
\end{figure*}

\subsubsection{UV-to-NIR SED fitting with \texttt{CIGALE} at z$_{spec}$}
\label{subsec: CIGALE}



Unlike SED fitting codes, which fit only the UV to NIR part of the SED (e.g., FAST$++$) and can suffer from degeneracies between the age, the metallicity, and the attenuation of a galaxy, an alternative approach is gaining popularity. It relies on the energy balance principle, which states that the energy emitted by dust in the mid- and far-IR exactly matches the energy absorbed by dust in the UV-optical range. This approach has been adopted and implemented by codes such as \texttt{CIGALE} \citep{Boquien2019}, \texttt{MAGPHYS} \citep{DaCunha2008, DaCunha2015}, \texttt{BAGPIPES} \citep{Carnall2018}, and \texttt{PROSPECTOR} \citep{Johnson2021}.



In this section, we utilize \texttt{CIGALE} (Code Investigating GALaxy Emission)\footnote{CIGALE v2022.1: \url{https://cigale.lam.fr/}} \citep{Boquien2019, Yang2020, Yang2022}, a popular SED fitting code that models galaxy emission from the UV to radio wavelengths using the energy balance principle.
Compared to FAST$++$, \texttt{CIGALE} provides constraints on reliable obscured star formation and AGN contributions across the full UV-to-radio SED.
We use six modules (sfhdelayed, bc03, nebular, dustatt\_modified\_starburst, dl2014, skirtor2016)
and the same UV-to-NIR photometry utilized in Sections\ \ref{subsec:FAST} and\ \ref{subsec:FAST++}.
Below, we briefly summarize the parameters we adopted. Full details may be found in Table\ \ref{tab:CIGALE_para}. 
We utilize a delayed SFH model and simple stellar population (SSP) model from \citet{BruzualCharlot03}, along with a \citet{Chabrier2003} IMF. Nebular emission is included using the ionization parameter \( U \) of [-3, -2, -1] and gas metallicity \( Z_{\rm gas} \) values of [0.004, 0.008, 0.02, 0.05] as specified by \citet{VillaVelez2021}. 
In order to maintain consistency with the FAST$++$ fitting carried out in \citet{Forrest2020b}, we apply the modified dust attenuation model from \citet{Calzetti2000}. We adopt the dust emission model from \citet{Draine2014}. 
The updated \texttt{SKIRTOR} clumpy torus model \citep{Stalevski2012, Stalevski2016} is utilized to estimate the emission from an AGN disk and torus. 
The resulting Bayesian-estimated properties are shown in the fourth column of Tables~\ref{tab:all_DR3} and~\ref{tab:all_DR1}.

\subsubsection{UV-to-radio SED fitting with \texttt{CIGALE} at z$_{spec}$}
\label{subsec:CIGALEradio}

When utilizing the full UV-to-radio photometry with additional ten FIR-to-radio passbands in Table\ \ref{tab:passband}, we employ an extra ``radio" module in \citep{Boquien2019} to fit non-thermal synchrotron emission associated with star formation.
This relies on the radio-IR correlation ($\mathrm{q}_{\rm IR}$) and a power-law radio spectrum with the index $\alpha =$ 0.7. AGN radio emission is modeled via a component parameterized by the radio-loudness (\mbox{$R_{\mathrm{AGN}} = L_{\nu,\,5\,\mathrm{GHz}}/L_{\nu,\,2500\,\text{\AA}}$}), where \mbox{\( L_{\nu,\,\mathrm{5\,GHz}} \) and \( L_{\nu,\,2500\,\text{\AA}} \)} are the monochromatic AGN luminosities per unit frequency at RF 5 GHz and 2500 \text{\AA}, respectively \citep{Yang2022}.
The resulting estimates of physical properties are shown in the last (fifth) column of Tables~\ref{tab:all_DR3} and~\ref{tab:all_DR1}. 
Both UMGs are found to have AGN contributions across the whole wavelength, with AGN fractions of approximately 0.60 for \UMGB\ and 0.31 for \UMGA.

Figure~\ref{fig:CIGALE_two} shows the best-fit SED (black, upper panel) and residuals (lower panel) for \UMGB\ (left) and \UMGA\ (right), as estimated by \texttt{CIGALE} from the UV-to-radio photometry.
Also shown are the unattenuated stellar (blue), dust (red), AGN (cyan; including emission from torus, accretion disk, and polar dust), and radio synchrotron non-thermal (orange) emission. The multi-passband photometry is shown by purple circles, with downward arrows indicating upper limits where uncertainties exceeded the fluxes. 
We discuss the differences in the estimates of stellar mass (SM) and SFR in Section\ \ref{ssec:SMSFR}, and the difference in AGN fraction in Section\ \ref{subsec: AGN_frac}.

In order to explore the effect of adopting different dust attenuation laws in fitting the UV-to-radio SED, we compared the results of the \citet{Calzetti2000} model with those of the \citet{CharlotFall2000} model.
For the dusty \UMGA, the \citet{CharlotFall2000} attenuation model resulted in an estimate of stellar mass 0.5 dex higher than that obtained with the starburst \citet{Calzetti2000} model. In contrast, for the less dusty \UMGB, the difference in stellar properties was negligible. This comparison highlights the importance of the choice of dust attenuation model on the stellar mass estimate in the case of dusty galaxies. 
We defer a full investigation of this point to future work with a larger sample of UMGs.


\begin{figure*}
\centering
\includegraphics[width = 0.45\textwidth]{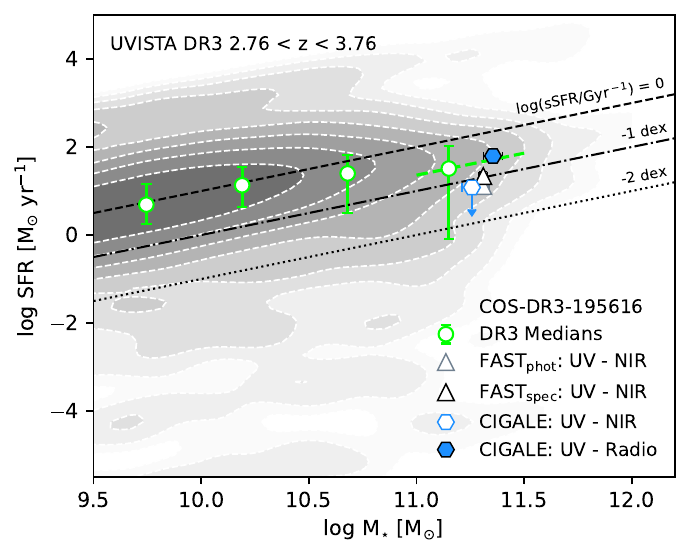}
\includegraphics[width = 0.45\textwidth]{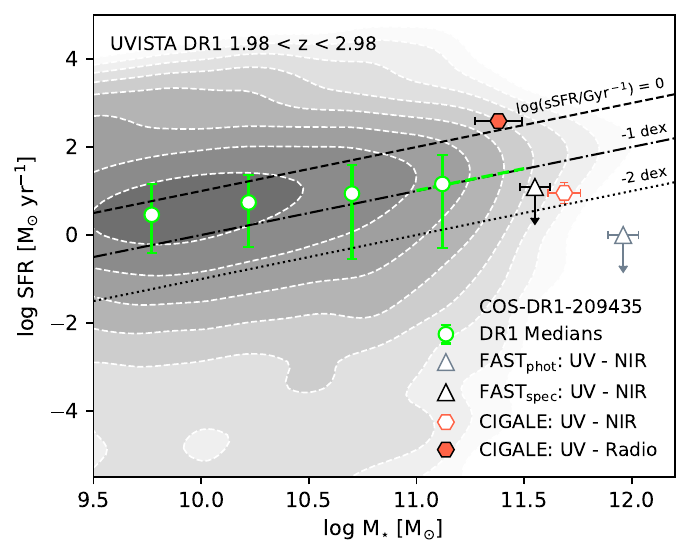}
\caption{
The SFR-stellar mass plane showing COS-DR3-195616 (left) and COS-DR1-209435 (right). The population of galaxies in the UltraVISTA DR3 (DR1) catalog within a redshift range of $z_{spec}\pm 0.5$ of each UMG is indicated by the gray contours. 
The green symbols show the median SFR in bins of 0.5 dex in stellar mass, with error bars representing the 16th and 84th percentiles of the distribution of SFR among galaxies in each bin.  
The median sSFR of the highest mass bin is shown as a dashed green line with $\log$(sSFR/Gyr$^{\rm -1}$) $\sim-0.64$ ($-0.99$) for the DR3 (DR1) population.  
The open gray and black triangles show the FAST$++$ fits the UV-NIR SEDs at the photometric and spectroscopic redshifts, respectively. The open and filled blue and red hexagons show the \texttt{CIGALE} fits the UV-NIR and UV-radio SEDs, respectively, at the spectroscopic redshifts (see Section\ \ref{sec:analysis} and Tables\ \ref{tab:all_DR3} and ~\ref{tab:all_DR1}).
}
\label{fig:SFMS}
\end{figure*}

\subsection{Stellar Mass and Star Formation Rate} 
\label{ssec:SMSFR}

SM and SFR are two of the most important properties that can be measured for any galaxy.
In Figure~\ref{fig:SFMS}, we compare the four different estimates of SM and SFR which we obtained for \UMGB\ and \UMGA\ in Section\ \ref{ssec:analysis}.
In each panel, the open gray and black triangles show the FAST$++$ fits to the UV-NIR SEDs at the photometric and spectroscopic redshifts, while the open and filled blue/red hexagons show the \texttt{CIGALE} fits to the UV-NIR and UV-radio data at the spectroscopic redshifts (which correspond to the values in columns 2-5 in Tables~\ref{tab:all_DR3} and~\ref{tab:all_DR1}).
We also compare these values with the populations of galaxies in the UltraVISTA DR1\&DR3 catalog with \mbox{$|z_{\rm phot} - z_{\rm spec, UMG}| < 0.5$} for each UMG, indicated by the gray contours.
These contours are generated using a 2D Gaussian kernel density estimation (KDE). 
The median values in 0.5 dex bins of stellar mass and 16th/84th percentiles of the SFR distribution are shown in green.
The dashed green line shows the star-forming main sequence (SFMS) for the highest mass bin, represented by the median sSFR of galaxies with \logM\ $> 11$.


In the case of \UMGB\ (Figure~\ref{fig:SFMS} left), the spectroscopic redshift ($z_{\rm spec}=3.255$) is similar to the photometric redshift ($z_{\rm phot}=3.09$) and therefore, unsurprisingly, there is little difference between the SM and SFR estimates derived by FAST$++$ from the UV-to-NIR photometry at $z_{\rm phot}$ and $z_{\rm spec}$ (columns 2 and 3 of Table~\ref{tab:all_DR3}; open gray and black triangles in the left panel of Figure~\ref{fig:SFMS}).
The SM estimates obtained using \texttt{CIGALE} are consistent within 0.1 dex with those from FAST$++$, when using either the same UV-to-NIR photometry (open blue hexagon) or the UV-to-radio photometry (filled blue hexagon). 
The \texttt{CIGALE} SFR estimate using the UV-to-NIR photometry is also similar to that from FAST$++$.
However, the \texttt{CIGALE} SFR estimate using the UV-to-radio photometry results in a somewhat ($\sim$3-5$\times$) higher SFR estimate of \mbox{$62.5\pm9.6$ $M_{\odot}$yr$^{-1}$}, consistent with the SFMS value for the highest mass bin derived from the UltraVISTA catalog. 
This increase reflects the contribution of mildly obscured star formation, consistent with its modest dust attenuation (\mbox{$A_{V} = 0.68\pm0.09$}).


However, in the case of \UMGA\ (right panel in Figure~\ref{fig:SFMS}), 
the various methods result in notable differences in the SM and SFR estimates.
The spectroscopic redshift ($z_{\rm spec}=2.48$) differs significantly from the photometric estimate ($z_{\rm phot}=3.76$), and this result in significant differences in the FAST$++$ SM and SFR estimates derived from the UV-to-NIR photometry (columns 2 and 3 of Table~\ref{tab:all_DR1}; open gray and black triangles in the right panel of Figure~\ref{fig:SFMS}), and in the \texttt{CIGALE} SM and SFR estimates derived from the UV-to-NIR and UV-to-radio photometry (columns 4 and 5 of Table~\ref{tab:all_DR1}; open and filled red hexagons in the right panel of Figure~\ref{fig:SFMS}).
Focusing on the four SM estimates first, we note that while the differences among them are larger than those for \UMGB, the largest contributing factor is caused by the offset between the photometric and spectroscopic redshift. When adopting only the spectroscopic redshift, the estimates of SM from the different photometry/fitting codes are in agreement within 0.3 dex (columns 3-5). 
The story becomes more interesting when we turn to the SFR. 
Analysis of the UV-to-NIR photometry suggests that \UMGA\ is forming stars at a very low rate of less than \mbox{14 $M_{\odot}$ yr$^{-1}$}, implying that the galaxy is in the final stages of shutting down star formation (see columns 2–4 in Table~\ref{tab:all_DR1}).
However, the \texttt{CIGALE} estimate of SFR derived from the full UV-to-radio SED (column 5) paints a much different picture, revealing that \UMGA\ is actually undergoing intense star formation at a rate of \mbox{$389.1 \pm 83.4$ $M_{\odot}$yr$^{-1}$} with heavy dust obscuration ($A_V = 2.92 \pm 0.19$). 
The dust-to-stellar mass ratio of \UMGA\ ($M_{\rm dust}/M_{\star}$ = 10$^{-1.7}$) is slightly elevated compared to typical main-sequence galaxies at the similar mass and redshift ($M_{\rm dust}/M_{\star}\sim10^{-2}$; \citealt{Scoville2017, Tacconi2018}), further supporting its classification as a DSFG.

As a sanity check, we also estimated parameters using modified \texttt{MAGPHYS} with ``high-$z$'' expansion \citep{DaCunha2015}, including the AGN template (\citealt{Chang2017}; da Cunha \& Battisti et al.\ in prep.). 
We obtained similar estimates for SM and SFR as when we utilized \texttt{CIGALE}.
We conclude that the very elevated SFR for \UMGA\ derived from \texttt{CIGALE} when including FIR-to-radio data is not due to systematic differences between SED fitting codes, but rather underscores the importance of FIR observations to accurately recover hidden star formation in dusty galaxies. 
However, in contrast to \texttt{CIGALE}, \texttt{MAGPHYS} lacks a radio module and does not consider AGN radio emission in the SED fitting. Consequently, 
in the following sections, we use the SFR and SM estimates obtained from the UV-to-radio SED fits with \texttt{CIGALE} to investigate additional properties of our galaxies.


\subsection{Molecular Gas Mass and Depletion Timescale}
\label{subsec: ALMA}
 
FIR observations, which trace the full dust emission, allow us to estimate molecular gas masses and gas depletion timescales. These quantities are essential for understanding the molecular gas reservoir and the mechanisms that fuel or quench star formation. 
In the following section, we present estimates of molecular gas mass and gas depletion timescale for each UMG and then explore implications for their evolution.
 

\subsubsection{Molecular Gas Mass}
\label{subsubsec: dis_Mgas}

\input{Table_Gas_RJ_Mdust}
  
Several studies have shown that the molecular gas masses of galaxies correlate tightly with the dust continuum luminosity in the Rayleigh-Jeans (RJ) tail of their SED (e.g., \citealt{Scoville2014, Scoville2016}; \citealt{Groves2015}; \citealt{Hughes2017}). In this work, we adopt the empirically derived luminosity-to-mass ratio ($\alpha_\nu$) from \cite{Scoville2016}: 
\begin{equation}
\label{eq: ratio_L_M}
\alpha_\nu \equiv \left\langle \frac{L_{\nu_{850\mu m}}}{M_{\text{mol}}} \right\rangle = 6.7 \pm 1.7 \times 10^{19} \, \text{erg} \, \text{s}^{-1} \, \text{Hz}^{-1} \, M_\odot^{-1} .
\end{equation}

It is common practice to scale ALMA observations to RF 850\um\ and use equation~(\ref{eq: ratio_L_M}) to estimate the molecular gas mass when the observation probes the RJ tail (\mbox{$\lambda_{\rm rest}>250$ \um}). However, at the redshifts of these UMGs, Band 7 (872 \um) corresponds to \mbox{$\lambda_{\rm rest}=206.4$ \um} for \UMGB\ and \mbox{$\lambda_{\rm rest}=250.4$ \um}
for \UMGA. 
Instead, we estimated the RF 850\um\ luminosity ($L_{\nu_{850,\mu{\rm m}}}$) directly from the \texttt{CIGALE} fit.
To account for differences between the \texttt{CIGALE}-derived dust temperatures and the value of 25K utilized in \citet{Scoville2016} calibration, we applied a correction to the RF 850\um\ luminosity using the ratio, $B_{\nu}(850, 25K)/B_{\nu}(850, T_{\text{dust}}$), where $B_{\nu}(\lambda, T_{\text{dust}})$ is the Planck function and T$_{\text{dust}}$ is derived from \texttt{CIGALE} using the model in \citet{Draine2014}.  
This correction factor is 0.77 for \UMGB\ and  1.06 for \UMGA. 
Molecular gas mass estimates based on \citet{Scoville2016} are shown in Table~\ref{tab: Mgas_tdep}. 
 
For comparison, we also estimate the molecular gas mass in two additional ways. Firstly, we utilize the dust continuum method once again, but adopt the \citet{Hughes2017} calibration, which also relies on 850\um\ luminosity, but is based on a different empirical relation than \citet{Scoville2016}.  
The resulting molecular gas mass estimates differ by less than 0.15 dex.
Secondly, we adopt the dust mass obtained from the dust emissivity model \citep{Draine2014} in the \texttt{CIGALE} SED fits and estimate the molecular gas mass based on the gas-to-dust ratio. The estimate carry uncertainties due to assumptions about both the gas-to-dust ratio and the molecular gas fraction. Further details of the dust-mass derivation are provided in Appendix \ref{app: dust mass}. 
Table~\ref{tab: Mgas_tdep} presents the molecular gas mass (and depletion timescale) estimates derived from these two additional methods, showing consistency within 0.2 dex. 
Therefore, we utilize the \citet{Scoville2016} values, hereafter. 

\begin{figure}
\centering 
\includegraphics[width = 0.45\textwidth]{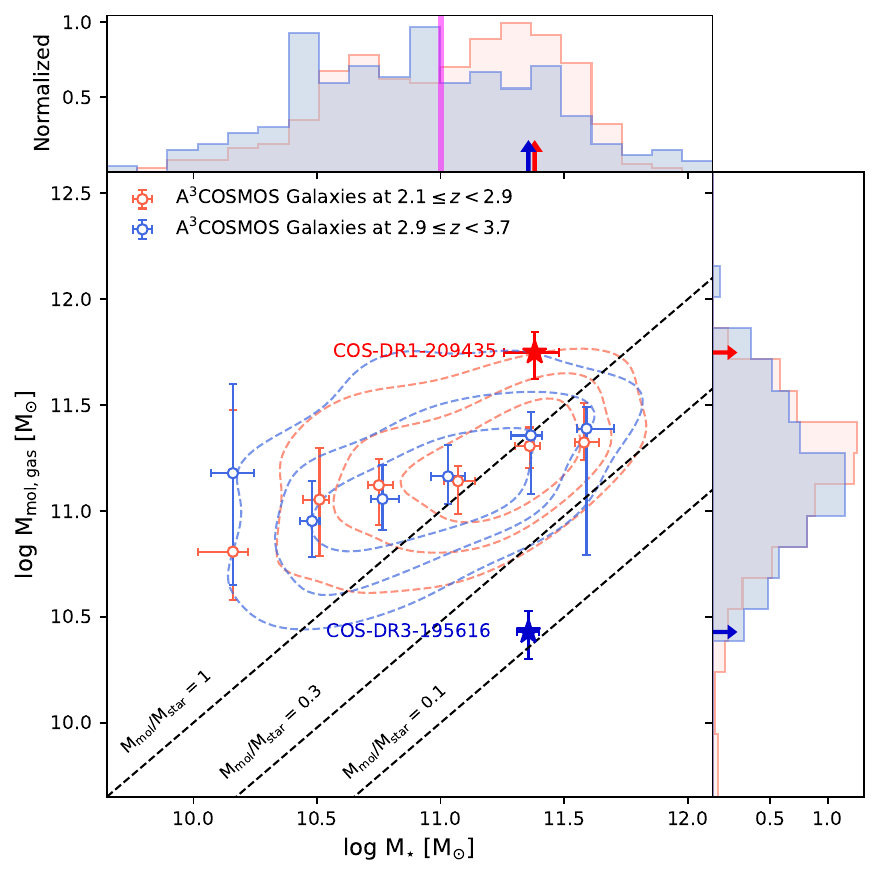}
\caption{
Molecular gas mass versus stellar mass with \UMGB\  (\UMGA) shown by the solid blue (red) star.
The top and right panels show the normalized histogram of stellar mass and molecular gas mass, respectively, for A$^{3}$COSMOS galaxies in the redshift range $2.1 \leq z < 2.9$ (red) and $2.9 \leq z < 3.7$ (blue).
The colored contours enclose 25\%, 50\%, and 75\% of the galaxy population, showing the two-dimensional density distributions for A$^{3}$COSMOS catalog in each redshift range, constructed using kernel density estimation.
The open colored circles show medians, with 3$\sigma$ bootstrap uncertainties. 
The dashed black lines show gas-to-stellar mass ratios of 0.1, 0.3 and 1.
The magenta vertical line in the top panel shows the lower mass limit (\mbox{\logM\ $> 11$}) used to make Fig.\ \ref{fig:Mgas_Msfr_two}.  
}
\label{fig:Mgas_Mstar_two}
\end{figure}

Figure~\ref{fig:Mgas_Mstar_two} shows molecular gas mass 
versus stellar mass with
\UMGB\ shown by the solid blue star and \UMGA\ by the solid red star. 
The top and right panels show the normalized histograms of stellar mass and molecular gas mass, respectively,
for A$^{3}$COSMOS (ALMA-detected) galaxies with a spectroscopic redshift of $z_{\mathrm{spec}} \pm 0.4$ of each UMG's spectroscopic redshift (see also Table~\ref{table: A3COSMOS}).
The colored contours enclose 25\%, 50\%, and 75\% of the galaxy population, showing the two-dimensional density distribution for A$^{3}$COSMOS galaxies in the redshift range $2.1 \leq z < 2.9$ (red) and $2.9 \leq z < 3.7$ (blue), constructed using a KDE to smooth the invididual data points and highlight regions of higher and lower source density.   
For consistency, we adopt the A$^{3}$COSMOS molecular gas masses from \citet{Liu2019b}, derived from dust continuum observations using the method of \citet{Scoville2016}. However, we note that the \citet{Liu2019b} approach does not include any possible AGN contribution, which may lead to overestimates of both SFR and molecular gas mass.   
The open colored circles show medians with 3$\sigma$ bootstrap uncertainties. 
The dashed black lines show gas-to-stellar mass ratios of 0.1, 0.3, and 1.  
 
Within the scatter, there does not appear to be any significant difference between the molecular mass-stellar mass relationship between the two redshift ranges.
At the high mass end, both populations appear to lie close to the \mbox{\(M_{\rm mol}/M_{\star} = 1\)} relationship.  
The dusty star-forming galaxy, \UMGA\ (red star), has a high molecular gas-to-stellar mass ratio (\mbox{\(M_{\rm mol}/M_{\star} = 2.33 \pm 0.83\)}). 
It has one of the largest molecular gas reservoirs of all of the galaxies in the A$^{3}$COSMOS catalog at the similar redshifts.  
In contrast, \UMGB\ (blue star) has a very low gas-to-stellar mass ratio (\mbox{\(M_{\rm mol}/M_{\star} = 0.12 \pm 0.03\)}), making it
one of the smallest molecular gas reservoirs of all of the galaxies in the A$^{3}$COSMOS catalog at a similar redshift.
We shall return to discuss possible reasons for this small molecular mass reservoir in Section\ \ref{subsec: AGN_role}.

\begin{figure*}
\centering
\includegraphics[width = 0.45\textwidth]{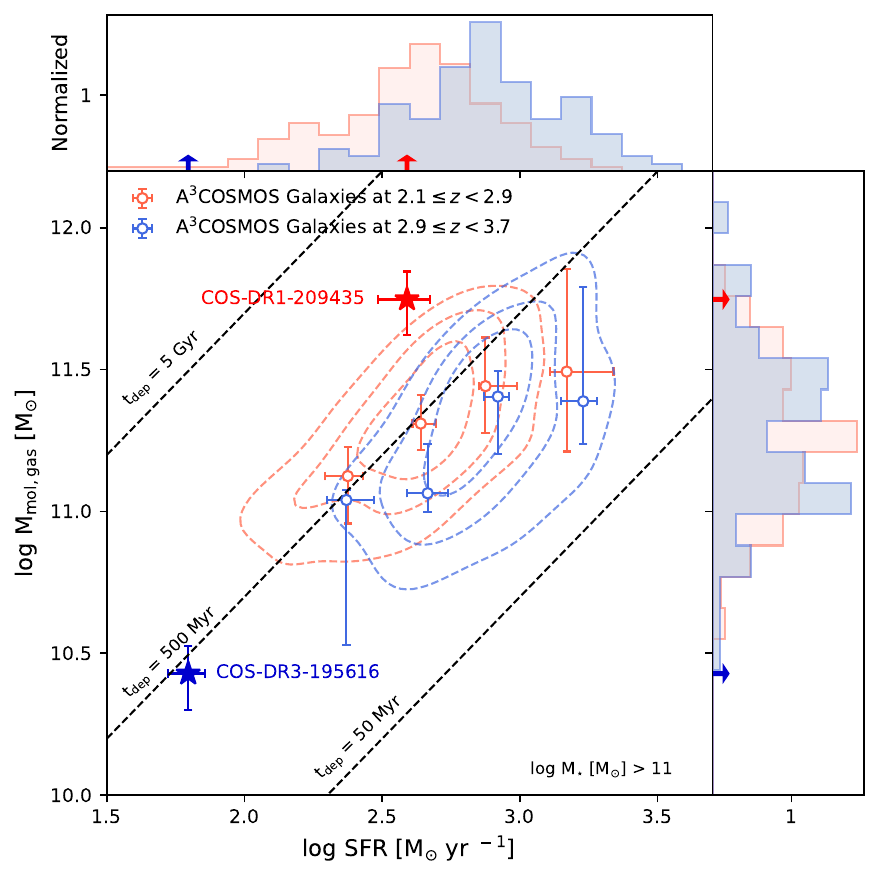}  
\includegraphics[width = 0.45\textwidth]{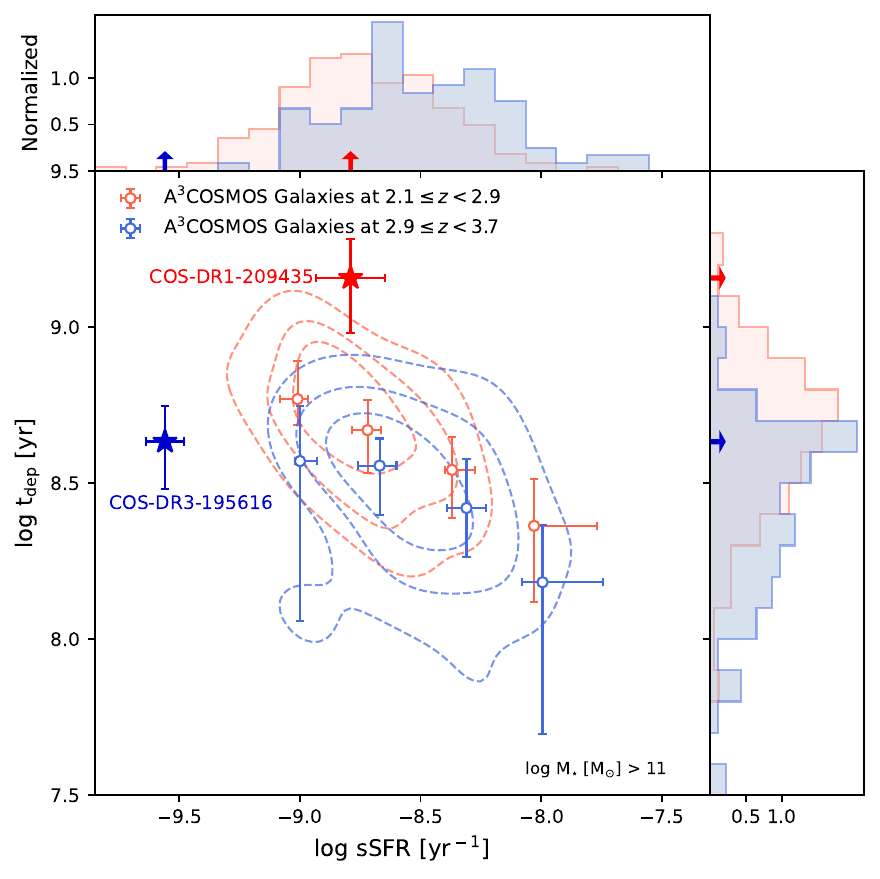} 
\caption{ 
Molecular gas mass as a function of star formation rate (left) and depletion timescale as a function of specific star formation rate (right). The symbols are as in Figure~\ref{fig:Mgas_Mstar_two}, but note that only A$^{3}$COSMOS galaxies with a stellar mass of \logM$>11$ were used to make this figure (see Table~\ref{table: A3COSMOS}).
The dashed black lines in the left panel correspond to gas depletion timescales of 50 Myr, 500 Myr, and 5 Gyr. 
}
\label{fig:Mgas_Msfr_two}
\end{figure*}

\begin{deluxetable}{lcc} 
\setlength{\tabcolsep}{26pt}
\tablecaption{Number of A$^{3}$COSMOS Galaxies (Fig.~\ref{fig:Mgas_Mstar_two}) and Number above the SM limit of \logM\ $> 11$ (Fig.~\ref{fig:Mgas_Msfr_two})
as a function of redshift range.
\label{table: A3COSMOS}
}
\tablehead{
\colhead{Redshift} & \colhead{Total} & \colhead{UMGs}
}
\startdata
$2.1 \leq z < 2.9$ & 315 & 174 \\
$2.9 \leq z < 3.7$ & 224 & 93 \\
\enddata
\end{deluxetable}

\subsubsection{Depletion Timescale}
\label{subsubsec: T_dep}

\input{Table_Prop_in_figures}

Table\ \ref{tab:Figure_prop} summarizes the stellar and gas properties of the two UMGs discussed in the following section. 
Figure~\ref{fig:Mgas_Msfr_two} shows molecular gas mass as a function of star formation rate (left) and gas depletion timescale (\mbox{$t_{\rm dep} =M_{\rm mol}/$SFR}) as a function of specific star formation rate (right). The symbols are as in Figure~\ref{fig:Mgas_Mstar_two}. However, to minimize the effect of stellar mass, we limited the analysis to A$^{3}$COSMOS galaxies with a stellar mass of \mbox{\logM\ $> 11$} (magenta line in Figure~\ref{fig:Mgas_Mstar_two}) in creating Figure~\ref{fig:Mgas_Msfr_two}.
As can be seen from Table~\ref{table: A3COSMOS}, approximately half of all of the A$^{3}$COSMOS galaxies in each of the two redshift ranges have a stellar mass of \mbox{\logM\ $> 11$}.
The dashed lines in the left panel of Figure~\ref{fig:Mgas_Msfr_two} correspond to gas depletion timescales of 50 Myr, 500 Myr, and 5 Gyr, indicating the time it would take for galaxies to consume all of their molecular gas if they continued to form stars at their current SFR. 

In Figure~\ref{fig:Mgas_Msfr_two}, both \UMGB\ and \UMGA\ stand out as outliers on both the $M_{\rm mol}-$SFR diagram (left panel) and $t_{\rm dep}-$sSFR diagram (right panel).
The dusty star-forming UMG, \UMGA, despite having one of the largest molecular gas reservoirs among its peers, \mbox{$\log (M_{\rm mol}/M_\odot) = 11.75\pm0.11$}, has a relatively typical SFR. This results in a long gas depletion timescale of more than 1 Gyr (\mbox{$1.44\pm0.48$ Gyr}), indicating a relatively low star formation efficiency. 
In contrast, \UMGB\ has both a small molecular gas mass of \mbox{ $\log (M_{\rm mol}/M_\odot) = 10.43\pm0.11$} and a low SFR relative to A$^{3}$COSMOS galaxies of similar stellar mass and redshift.
Its depletion timescale, $t_{\rm dep} =0.43\pm 0.13$ Gyr, is similar to other A$^{3}$COSMOS galaxies of similar stellar mass and redshift (right panel),
suggesting that \UMGB\ is forming stars with relatively typical efficiency, despite its limited fuel supply.  
We note that \UMGB\ exhibits modest AGN contribution in the FIR (see Section\ \ref{subsec: AGN_role}), which may affect molecular gas mass estimates, particularly when using the RJ tail calibration from \citet{Scoville2016}. However, this potential contribution from AGN could lead to an overestimate of the gas mass, which would imply that the true molecular gas mass for \UMGB\ is even lower and further strengthens the case for it being an outlier compared to other A$^{3}$COSMOS galaxies in Figure\ \ref{fig:Mgas_Mstar_two} and Figure\ \ref{fig:Mgas_Msfr_two}. 

\subsection{Predicted Stellar Mass Evolution }
\label{subsec: Mass_evolution}

\begin{figure*}
\centering 
\includegraphics[width = 0.98\textwidth]{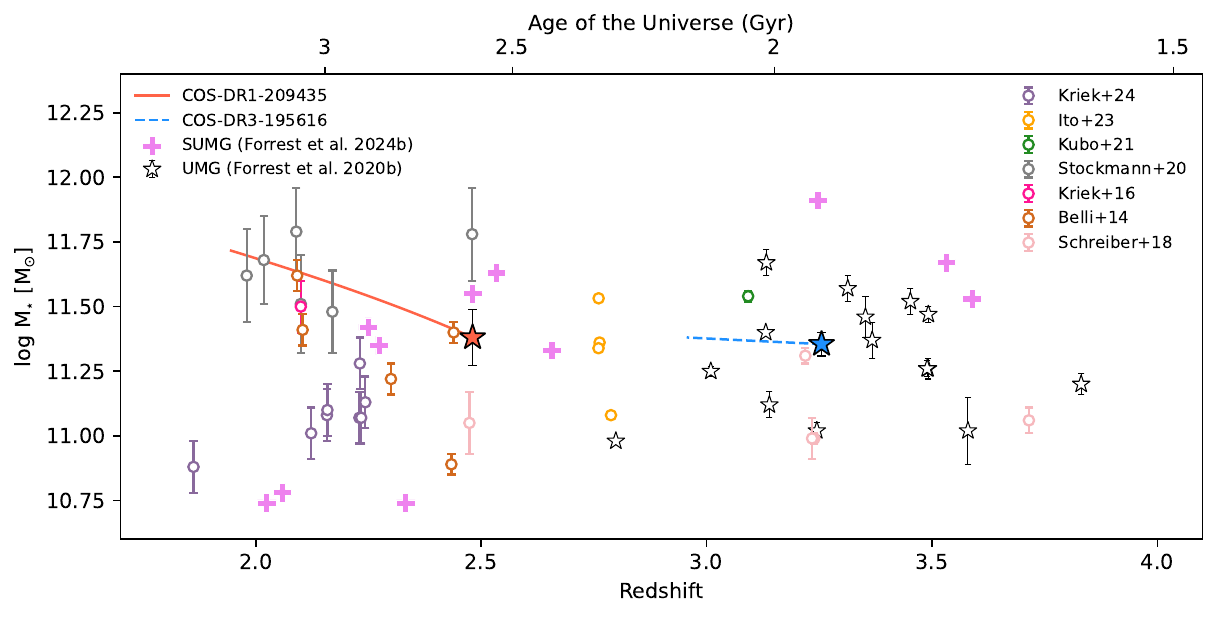} 
\caption{Stellar mass evolution predictions for \UMGBl\ (blue) and \UMGAl\ (red) assuming a constant SFR for one half of the gas depletion timescale. Black empty stars and violet crosses are as in Figure~\ref{fig: UVJ}.
COS-DR1-209435 is anticipated to grow in stellar mass to \logM\ = 11.72 within 0.72 Gyr and COS-DR3-195616 to \logM\ = 11.38 within 0.21 Gyr. Open circles show spectroscopically confirmed quiescent massive galaxies 
from the literature.  
\label{fig:Mass_evol} }
\end{figure*}

In order to predict the stellar mass evolution of the UMGs, we assume a constant SFR over one half of the depletion timescale (0.5$\times t_{\rm dep}$).
As shown in Figure~\ref{fig:Mass_evol}, the significantly different molecular gas mass and SFR properties of the two UMGs result in significantly different stellar mass evolutionary pathways.
Having already depleted most of its molecular gas, and with a relatively low SFR of \mbox{$62.5\pm9.6$ $M_{\odot}$yr$^{-1}$}, \UMGB\ is anticipated to increase only 0.02 dex in stellar mass and reach a stellar mass of \mbox{\logM\ = 11.38} over the next 0.21~Gyr. In contrast, \UMGA\ is highly gas-rich and still actively forming stars at a SFR of \mbox{$389.1\pm83.4$ $M_{\odot}$yr$^{-1}$}. It is anticipated to increase 0.34 dex in stellar mass, reaching a stellar mass of \mbox{\logM\ = 11.72} over the course of the next 0.72~Gyr. 
\UMGB\ is expected to grow by only $\sim5\%$ in stellar mass, suggesting it appears to be approaching the end of its stellar mass assembly.
In contrast, \UMGA\ is projected to more than double its current stellar mass, with an anticipated increase of $\sim120\%$, indicating it is still in an active phase of stellar mass growth.



\section{Discussion}
\label{sec: Discussion}



\subsection{AGN Diagnostics}
\label{subsec: AGN_frac}

\citet{Marsan2017} found that a significant fraction of $3 < z < 4$ UMGs host powerful AGN, motivating us to assess potential AGN existence in our two UMGs.
We analyze a number of AGN diagnostics, as we describe below and summarize in Table~\ref{tab: AGN_classification}.

\subsubsection{AGN Fraction and Bolometric Luminosity}
 
In this work, we utilized the \texttt{SKIRTOR} module in \texttt{CIGALE} to estimate the AGN fraction, frac$_{\rm AGN}$, 
defined as \mbox{$L_{\rm dust, AGN}/$($L_{\rm dust, AGN}$ + $L_{\rm dust, galaxy}$)}, where $L_{\rm dust, AGN}$ and $L_{\rm dust, galaxy}$ refer to the luminosity from dust heated by AGN and from star formation \citep{Yang2022}.
UMG \UMGB\ exhibits a relatively high AGN fraction (frac$_{\rm AGN}$ = 0.60 $\pm$ 0.03) across the RF 1-3000 \um, indicating that 60\% of the dust-emitted radiation is associated with AGN activity. In contrast, \UMGA\ has an AGN 
fraction of $0.31\pm0.10$. 
We further assess the AGN contribution in the mid-infrared (MIR; RF 1-30 $\mu$m), where warm dust heated by AGN or star formation emits strongly \citep{DraineLi2007}.
As shown in Table~\ref{tab: AGN_classification}, both UMGs exhibit strong MIR AGN fractions (\mbox{frac$_{\rm AGN, MIR} >$ 70\%}), suggesting that warm dust emission is primarily driven by AGN activity in two UMGs.
We also estimate the frac$_{\rm AGN}$ in the RJ tail (RF 250-1000$\mu$m), the wavelength range used to estimate dust mass and hence molecular gas mass by the \citet{Scoville2016} method. The frac$_{\rm AGN}$ for \UMGA\ is only 2\% in this regime, indicating negligible contamination, while \UMGB\ shows a frac$_{\rm AGN}$ of 18\%, suggesting a modest level of AGN influence on the FIR emission. We conclude, therefore, that the continuum emission in both galaxies originates primarily from thermal dust that is heated by the radiation from star formation. 
We estimate the Bayesian AGN bolometric luminosity ($L_{\rm bol, AGN}$) from the \texttt{CIGALE} SED fitting, which represents the total AGN emission across all wavelengths, including contributions from the accretion disk and dust-heated infrared emission, as listed in Table~\ref{tab: AGN_classification}. 
Both UMGs show $L_{\rm bol, AGN} > 10^{45}$ erg s$^{-1}$, placing them among the population of luminous AGNs where the AGN likely plays a significant energetic role in the host galaxy \citep{Juneau2013}.

\subsubsection{X-ray and Radio AGNs}

To further assess evidence for AGN activity, we search for counterparts in the Chandra COSMOS-Legacy catalog \citep{Civano2016}, identifying only \UMGB\ as an X-ray source within a search radius of $1\arcsec$, with a positional offset of $0.11\arcsec$ from the optical counterpart.
We use equation (4) from \citet{Marchesi2016} to estimate the RF $2-10$~keV luminosity as \mbox{$L_{\rm 2-10 keV, rest} = 2.17 \pm 0.23 \times 10^{44}$ erg s$^{-1}$}, assuming a photon index of $\Gamma = 1.4$ \citep{Marsan2022}. This exceeds the commonly adopted AGN threshold of \mbox{$L_{\rm 2-10 keV, rest}>10^{44}$ erg s$^{-1}$} for galaxies at $z >2$ \citep{Juneau2013, Lamastra2013}, leading us to classify \UMGB\ as an X-ray-identified AGN.
Assuming the Eddington-limit accretion, $L_{\rm bol}/L_{\rm Edd} = 1$, we derive a lower limit on the black hole mass of \mbox{$M_{\rm BH} \gtrsim 1.7 \times 10^7$ $M_\odot$}. If the AGN is sub-Eddington, the true mass could be significantly higher, suggesting the existence of a supermassive black hole in the center of \UMGB.


Both UMGs are detected at VLA 1.4 GHz (SNR $>$ 3$\sigma$) with the radio luminosities of \mbox{$L_{\text{1.4GHz}}=2.6 \pm 0.6 \times 10^{24}$ W Hz$^{-1}$} for \UMGB\ and \mbox{$L_{\text{1.4GHz}}=4.0 \pm 1.1 \times 10^{24}$ W Hz$^{-1}$} for \UMGA, derived using equation (7) from \citet{Butler2018} with a spectral index of $\alpha = -0.8$ \citep{Marsan2017}.
We classify a source as radio-loud AGNs if it meets either of the following criteria: (1) RF $L_{\rm 1.4GHz} > 10^{25}$ W Hz$^{-1}$ \citep{Butler2018}, or (2) radio-loudness parameter R$_{\rm AGN} > 10$ \citep{Yang2022}, derived from \texttt{CIGALE} SED fitting. 
Both UMGs fall below these thresholds and are thus classified as radio-quiet AGNs, where the emission is primarily radiative rather than jet-driven. As shown in Figure\ \ref{fig:CIGALE_two}, the observed radio fluxes are well fit by the SED model without excess indicative of a jet. The AGN component is prominent in the mid-IR, consistent with thermal dust from a torus, supporting the radio-quiet AGN classification.

\input{AGN_classification}
 
\subsubsection{Emission Lines}

Emission line ratios are also widely used to identify AGN activity \citep{Baldwin1981}. For \UMGB, we detected \OIIIlater\ and \Hbeta\ emission lines in the MOSFIRE $K$-band spectrum (Figure~\ref{fig: Ks_img_spectra}), with a line flux ratio of \mbox{$\log \mathrm{f_{5007}/f_{H\beta}}= 0.79\pm0.08$}, which is suggestive of potential AGN activity \citep{Wuyts2016, Forrest2020b}. 
At \UMGA's lower spectroscopic redshift, \OIIIlater\ and H$\beta$ fall outside the $K$-band, but \NIIlater\ and H$\alpha$ are detected with a flux ratio of \mbox{$\log \mathrm{f_{6584}/f_{H\alpha}}= -0.21\pm0.11$}. 
Considering the uncertainties, we cannot definitively rule out star formation as the dominant source of the nebular emission in \UMGA\ based on the \citet{Kewley2001b} diagnostic. 
Notably, \citet{Forrest2024b} modeled the blended \NIIlater\ and H$\alpha$ emission lines \UMGA, fitting a narrow H$\alpha$ component, finding no evidence for a broad H$\alpha$ component.
With the \OIIIlater\ line luminosities of \UMGB, we estimate $L_{\rm bol, AGN}= 1.70\pm0.99 \times 10^{45}$ erg s$^{-1}$ using the relation from \citet{Juneau2013}. 
This value is consistent within an order of magnitude with the SED-derived estimate in Table~\ref{tab: AGN_classification}, with the higher SED-based luminosity likely reflecting dust attenuation of the RF optical lines and the more comprehensive coverage of the SED. 


\subsection{UMG Nature and Evolutionary Path}
\label{subsec: AGN_role}

 
In this work, we studied two UMGs that capture distinct and critical stages of massive galaxy evolution, shaped by the interplay between gas, star formation, and AGN activity.
Our analysis reveals that \UMGB\ has a relatively suppressed specific star formation rate of \mbox{$\log(\rm sSFR) = -9.82$ yr$^{-1}$}.   
Its gas-to-stellar mass ratio of 12\% is significantly lower than that of A$^{3}$COSMOS galaxies of similar stellar mass and redshift, but remains higher than the typical ratio (about 1\%) found in local quiescent galaxies \citep{Lisenfeld2017}.
Moreover, \UMGB\ has a gas depletion timescale of $\sim$500 Myr, consistent with a typical star formation efficiency for star-forming galaxies at this epoch. 
These properties, along with its position in the quiescent wedge of the UVJ diagram, indicate that \UMGB\ is likely in the quenching process and transitioning toward quiescence at $z>3$, as its remaining gas reservoir is no longer capable of sustaining high star formation.
%
Notably, \UMGB\ exhibits a lower SFR than other ALMA-detected galaxies at similar redshift and stellar mass. Its RMS (0.13 mJy/beam) does not reflect an atypical observational depth, indicating that \UMGB\ was not detected due to an abnormally long exposure time.
\UMGB\ may thus represent a class of unobscured, low star-forming ALMA sources where AGN activity powers the infrared emission.
Our findings support a scenario where the AGN may have contributed to the ongoing quenching of \UMGB\ by ejecting or disrupting the cold gas reservoir, thereby depleting the fuel available for star formation. 
This mechanism is consistent with recent observational studies that identify AGN-driven outflows of neutral and ionized gas as a key factor in suppressing star formation during the transitional phase of massive galaxies at $z \gtrsim 2$ \citep{Belli2024, Davies2024, Bugiani2025}. 


In the case of \UMGA, we have a strikingly different story. Our analysis shows that it remains actively star-forming, with a moderately high SFR
and a substantial cold gas reservoir, exceeding that of galaxies at similar redshift and stellar mass. 
This leads to a longer depletion timescale compared to A$^{3}$COSMOS galaxies, indicating that its large gas reservoir is being converted into stars less efficiently.
These results suggest that \UMGA\ is still undergoing stellar mass assembly given its substantial gas reservoir, indicating an active growth stage, at an earlier evolutionary phase than \UMGB, and not yet transitioning toward quiescence.
Despite the lack of an X-ray detection, we identify the existence of AGN activity based on its high MIR AGN fraction (f$_{\rm AGN,MIR} = 89\%$) and elevated AGN bolometric luminosity inferred from SED fitting.
The significant dust extinction (\mbox{$A_V=2.92\pm0.19$}) indicates that the AGN is deeply embedded in dust and gas, which potentially absorbs X-ray emission while allowing reprocessed mid-IR and radio signals to escape. 
The coexistence of vigorous star formation and abundant cold gas indicates that \UMGA\ may be at an earlier evolutionary phase than \UMGB, where AGN feedback has not yet suppressed star formation but may begin to play a regulatory role as the UMG evolves. 
The large gas reservoir currently fuels star formation, but may also provide the conditions necessary for AGN-driven outflows or heating that could contribute to future quenching.

\subsection{Importance of FIR/Radio Observations in Determining Redshift, Dust, and SFR}


As we showed in Section\ \ref{ssec:SMSFR}, the addition of FIR/radio observations revealed that both UMGs have higher and better-constrained SFRs than would have been inferred from UV-to-NIR photometry alone.
This is particularly important for \UMGA, whose red SED led to an overestimated photometric redshift and underestimated dust content and SFR when using only UV-to-NIR data. The additional FIR/radio observations reveal that it is more dust-obscured and actively star-forming than previously inferred. But for the less dusty \UMGB, the inclusion of FIR photometry leads to only minor changes in estimated SFRs ($\Delta \log{\rm SFR} < 1$) and stellar masses ($\Delta$ \logM\ $< 0.1$).
Therefore, we are able to definitively classify \UMGA\ as a heavily dusty star-forming UMG and \UMGB\ as a less-dusty quenching (rather than quenched) UMG. 
Notably, the stellar mass estimates for two UMGs from the UV-NIR and UV-radio SED fitting agree within 0.3 dex, indicating that stellar mass measurements are robust against the inclusion of FIR data.


Therefore, our analysis demonstrates that relying solely on UV-to-NIR photometry risks systematically missing or misclassifying dust-obscured galaxies, potentially introducing significant biases in star formation rate estimates. 
By extending the wavelength coverage to the far-IR and radio, we gain access to the dust-obscured star formation and cold gas reservoirs that UV-NIR data cannot probe. This allows us to measure molecular gas masses, depletion timescales, and stellar mass growth, providing key insights into the gas that fuels star formation and regulates galaxy evolution. 
Beyond tracing star formation, these longer-wavelength data also enable a more complete quantification of AGN activity. As shown in Section~\ref{subsec:CIGALEradio}, both UMGs are 
characterized by strong mid- to far-IR emission and significant obscuration at shorter wavelengths. Without FIR data, the AGN contribution and the full energy output of these high-redshift galaxies would have been missed. 
Overall, our work highlights the essential role of far-IR and radio observations in fully characterizing the star formation, gas content, and AGN activity of massive galaxies, particularly during the dust-obscured phases of their evolution in the early Universe.

\section{Conclusions}
\label{sec: conc}

We matched publicly available ALMA (and other long-wavelength) observations from the A$^{3}$COSMOS catalog
to two spectroscopically-confirmed UMGs from the MAGAZ3NE survey, \UMGBl\ ($z_{\rm spec} = 3.255$) and \UMGAl\ ($z_{\rm spec} = 2.481$). 
By adding ten FIR-to-radio passbands to the existing UV-to-NIR MAGAZ3NE catalogs, we created new UV-to-radio photometric catalogs, which enabled a comprehensive, multi-wavelength analysis of the obscured star formation, gas content, and AGN activity in these two high-redshift galaxies. Our findings were as follows:

\begin{enumerate}

\item We explored how three factors, spectroscopic versus photometric redshift, choice of SED fitting code, and wavelength coverage (UV-NIR versus UV-to-radio), affected the key properties of the two UMGs.  
For \UMGB, whose photometric and spectroscopic redshifts are similar, we found that the SM estimates remained stable when derived by FAST$++$ or \texttt{CIGALE} using either the UV-to-NIR or the UV-to-radio photometry. However, the addition of FIR-to-radio data revealed a higher SFR, within 1 dex of the UV-NIR-only estimate, reaching \mbox{$62.5\pm9.6$ $M_{\odot}$yr$^{-1}$}. 
In contrast, for \UMGA, the photometric redshift was significantly overestimated relative to the spectroscopic redshift. Utilizing only the spectroscopic redshift, the estimates of SM from the different photometry/fitting codes were in agreement within 0.3 dex, but the SFR changed dramatically. The UV-NIR analysis (using either FAST$++$ or \texttt{CIGALE}) suggested it was a quiescent galaxy with \mbox{SFR $\leq 14 M_{\odot}$yr$^{-1}$}, while the full UV-radio SED fit revealed a much higher SFR of \mbox{$389.1\pm83.4$ $M_{\odot}$yr$^{-1}$}. 

\item Our analysis showed that the difference in SFR estimates from UV-NIR and UV-to-radio data was primarily driven by dust attenuation. UMG \UMGA, located in the upper-right region of the \textit{UVJ} diagram, showed substantial dust obscuration, 
which led to an UV-NIR underestimate of its star formation and contributed to its overestimated photometric redshift. By contrast, \UMGB, which lies in the quiescent wedge, had lower attenuation, 
allowing UV-NIR analyses to more accurately recover its star formation. 
The full UV-to-radio analysis revised our understanding of the nature of these two galaxies, revealing \UMGB\ to be an unobscured, quenching galaxy rather than a recently quenched UMG and  \UMGA\ to be a heavily obscured, actively star-forming system.
  
\item We estimated molecular gas masses using the RJ tail dust continuum measurement method. 
The quenching galaxy, \UMGB, was shown to be a gas-poor UMG, with one of the smallest molecular gas reservoirs and a much lower gas-to-stellar mass ratio among galaxies of similar mass and redshift. 
In contrast, the dusty star-forming UMG, \UMGA, exhibited one of the largest molecular gas reservoirs among its peers, reflected in its high molecular gas ratio and substantial cold gas supply available for star formation. 

\item Both \UMGB\ and \UMGA\ appear as outliers in $M_{\rm mol}-$SFR and $t_{\rm dep}-$sSFR diagrams.
UMG \UMGB\ has low gas mass and reduced SFR relative to A$^{3}$COSMOS peer galaxies, but showed a depletion timescale typical for A$^{3}$COSMOS galaxies of similar mass and redshift. 
This indicated that \UMGB\ was forming stars with relatively typical efficiency, despite its limited fuel supply.  
In contrast, \UMGA, despite hosting an exceptionally large molecular gas, has a relatively typical SFR amongst its peers, resulting in a prolonged gas depletion timescale of more than 1 Gyr, indicative of low star formation efficiency and stellar mass assembly. 

\item The two UMGs are predicted to follow different stellar mass growth trajectories, due to the contrast in their molecular gas reservoirs and SFRs.
UMG \UMGB\ has already depleted most of its molecular gas reservoir and is anticipated to experience minimal future stellar mass growth. In contrast, \UMGA\ is anticipated to increase 0.34 dex in stellar mass, reaching a stellar mass of \mbox{\logM\ = 11.72} over the course of the next 0.72~Gyr. 


\item  We presented multi-pronged evidence for significant AGN activity in both UMGs, revealed through a combination of X-ray, infrared, and emission line diagnostics, as well as high AGN fractions and bolometric luminosities from SED fitting.
In \UMGB, our findings supported a scenario where the AGN may have contributed to depleting the cold gas supply, potentially resulting in its quenching. 
Meanwhile, \UMGA\ continues to form stars despite hosting an obscured AGN, suggesting that AGN feedback has not yet suppressed star formation but may begin to play an important role as the galaxy evolves.

\end{enumerate}
 
This work reveals two UMGs captured at distinctively different evolutionary stages: one appears to be caught in the midst of rapid quenching, demonstrating AGN activity and with little remaining molecular gas; the other appears to be vigorously forming stars sustained by a rich gas reservoir.
Our work shows the importance of FIR-to-radio observations for accurately inferring the nature and properties of galaxies at $z\gtrsim3$. While UV-NIR observations alone can provide robust stellar mass estimates for less-obscured galaxies, they often fail to capture obscured star formation and hidden AGN activity, particularly in dusty systems. 
By adding FIR and radio photometry, we were able to recover obscured SFR, reveal the molecular gas reservoirs fueling SF and the growth in stellar mass, and quantify the AGN contribution across the full wavelength range, thus revealing the true pathways of UMG formation and quenching at early times.

\section{Acknowledgments}

The authors wish to recognize and acknowledge the very significant cultural role and reverence that the summit of Maunakea has always had within the indigenous Hawaiian community. We are most fortunate to have the opportunity to conduct observations from this mountain. 

GW gratefully acknowledges support from the National Science Foundation through grant AST-2347348.
BF gratefully acknowledges support from JWST-GO-02913.001-A.
AN gratefully acknowledges support from the National Science Foundation through grant AST-2307877, and from the Beus Center for Cosmic Foundations at Arizona State University.


\bibliographystyle{aasjournal}
\bibliography{sample631}

\appendix

\input{Table_CIGALE_input_t2}



\section{Molecular gas mass from Dust mass}
\label{app: dust mass}

 
Here, we describe the procedure we used to estimate molecular gas mass ($M_{\rm gas, mol}$) from dust mass ($M_{\rm dust}$).  \texttt{CIGALE} provides estimates of both the dust mass from the \citet{Draine2014} dust emission module and the gas-phase metallicity ($Z_{\rm gas}$) from the nebular emission module. We convert metallicity, $Z_{\rm gas}$, into oxygen abundances ($12 + \log(\mathrm{O/H})$) using the calibration from \cite{Asplund2009}. A more detailed discussion of oxygen abundance is provided in Appendix~\ref{subsec: metallicty}.
Our method follows a four-step process:  
(1) we estimate the gas-to-dust ratio (\mbox{$\delta_{\rm GDR} = M_{\rm gas, tot}/M_{\rm dust}$}; where \mbox{$M_{\rm gas, tot} = M_{\rm gas, mol}$ + $M_{\rm gas, atomic}$}) based on the oxygen abundance using the broken power-law relation in Table 1 of \citet{RemyRuyer2014}; (2) we calculate the total gas mass ($M_{\rm gas, tot} = M_{\rm dust} \times \delta_{\rm GDR}$), including both molecular and atomic components; (3) we estimate the fraction of molecular to total hydrogen ($f_{\rm mol}$) 
from oxygen abundance using the empirical relation from \citet{Krumholz2009}; (4) we derive the molecular gas mass as \mbox{\(M_{\rm gas,mol} = f_{\rm mol} \times M_{\rm gas, tot}\)}.

\section{Gas-phase Metallicity}
\label{subsec: metallicty}

\input{Table_OH_vertical}

We compare the SED-derived gas-phase metallicities with those predicted by established mass-metallicity relations (MZRs) and the fundamental metallicity relation (FMR) in Table\ \ref{tab: O_H_vertical}. 
Additionally, for \UMGA, the \NIIlater/\Halpha\ flux ratio is used to infer the gas-phase metallicity via the N2 index method \citep{PP04}. 
Our results show that the SED-derived gas metallicities agree well with the estimates from MZRs/FMRs and the N2 index, with a scatter of 0.1 dex. 
Compared to the MZR-based estimates, the CIGALE-derived oxygen abundance has a larger uncertainty, especially for \UMGB.
When estimating the molecular gas mass, we adopt the MZR-derived oxygen abundance from \citet{Maiolino2008} for \UMGB,
while for \UMGA, we instead adopt the oxygen abundance estimated using the N2 diagnostic.

\end{document}

%% file: Table_all_stellar_prop_DR3.tex
\begin{table}[htbp]
\centering
\caption{Comparison of the properties of UMG \UMGB\ estimated by fitting (left to right) the UV-to-NIR SED at $z_{\rm phot}=3.09$ using FAST$++$, the UV-to-NIR SED at $z_{\rm spec}=3.255$ using FAST$++$, the UV-to-NIR SED at $z_{\rm spec}$ using \texttt{CIGALE}, and the UV-to-radio SED at $z_{\rm spec}$ using \texttt{CIGALE}. 
\label{tab:all_DR3}}
\begin{tabular}{lllll}
\hline
\hline
Quantity & FAST$++$ & FAST$++$ &  CIGALE  & CIGALE  \\
  & \textnormal{\footnotesize UV--NIR} & \textnormal{\footnotesize UV--NIR} & \textnormal{\footnotesize UV--NIR} & \textnormal{\footnotesize UV--radio}\\
 & $z_{\rm phot}$ & $z_{\rm spec}$ & $z_{\rm spec}$ & $z_{\rm spec}$\\
\hline
$\log$ M$_{\star}$ & 11.31$^{\rm +0.03}_{\rm -0.03}$ & 11.31$^{\rm +0.02}_{\rm -0.02}$ & 11.26$^{\rm +0.04}_{\rm -0.05}$ & 11.36$^{\rm +0.04}_{\rm -0.05}$\\
SFR & 12.9$^{\rm +1.6}_{\rm -0.3}$ & 21.7$^{\rm +2.8}_{\rm -12.3}$ & $<$ 12.02 & 62.5 $^{\rm +9.6}_{\rm -9.6}$ \\
A$_{\rm v}$ & 0.6$^{\rm +0.1}_{\rm -0.1}$ & 0.6$^{\rm +0.1}_{\rm -0.2}$ & 0.16$^{\rm +0.12}_{\rm -0.12}$ & 0.68$^{\rm +0.09}_{\rm -0.09}$\\
frac$_{\rm AGN}$ & -- & -- & 0.38$^{\rm +0.22}_{\rm -0.22}$ & 0.60$^{\rm +0.03}_{\rm -0.03}$ \\
$\log$ Age & 8.93$^{\rm +0.01}_{-0.10}$ & 8.79$^{\rm +0.07}_{-0.01}$ & 9.00$^{\rm +0.02}_{\rm -0.02}$ & 9.00$^{\rm +0.01}_{\rm -0.01}$\\
\hline
\end{tabular}
\end{table}

%% file: Table_all_stellar_prop_DR1.tex
\begin{table}[htbp]
\centering
\caption{As for Table\ \ref{tab:all_DR3}, but for UMG \UMGA\ with $z_{\rm phot}=3.76$ and $z_{\rm spec}=2.481$. 
\label{tab:all_DR1}}
\begin{tabular}{lllll}
\hline
\hline
Quantity & FAST$++$ & FAST$++$ &  CIGALE  & CIGALE  \\
  & \textnormal{\footnotesize UV--NIR} & \textnormal{\footnotesize UV--NIR} & \textnormal{\footnotesize UV--NIR} & \textnormal{\footnotesize UV--radio}\\
 & $z_{\rm phot}$ & $z_{\rm spec}$ & $z_{\rm spec}$ & $z_{\rm spec}$\\
\hline
$\log$ M$_{\star}$ & 11.96$^{\rm +0.03}_{\rm -0.10}$ & 11.55$^{\rm +0.04}_{\rm -0.09}$ & 11.69$^{\rm +0.07}_{\rm -0.08}$ & 11.38$^{\rm +0.10}_{\rm -0.12}$\\
SFR & $<$ 1.0 & $<$ 12.3 & 9.0$^{\rm +5.0}_{\rm -5.0}$ & 389.1$^{\rm +83.4}_{\rm -83.4}$ \\
A$_{\rm v}$ & 0.9$^{\rm +0.1}_{\rm -0.1}$ & 2.9$^{\rm +0.1}_{\rm -0.3}$ & 2.00$^{\rm +0.24}_{\rm -0.24}$ & 2.92$^{\rm +0.19}_{\rm -0.19}$\\
frac$_{\rm AGN}$ & -- & -- & 0.10$^{\rm +0.01}_{\rm -0.01}$ & 0.31$^{\rm +0.10}_{\rm -0.10}$ \\
$\log$ Age & 9.20$^{\rm +0.01}_{\rm -0.18}$ & 8.63$^{\rm +0.18}_{\rm -0.19}$ & 9.33$^{\rm +0.08}_{\rm -0.09}$ & 9.28$^{\rm +0.09}_{\rm -0.10}$\\
\hline
\end{tabular}
\end{table}

%% file: Table_Gas_RJ_Mdust.tex
\begin{deluxetable*}{lcccccc}
\centering
\setlength{\tabcolsep}{10pt}
\caption{
Molecular gas mass and depletion timescales estimated from Rayleigh-Jeans tail dust continuum calibration methods of \citet{Scoville2016} and \citet{Hughes2017}
and gas-dust ratio ($\delta_{GDR}$) method. 
\label{tab: Mgas_tdep}}
\tablehead{
\colhead{UMG} & \colhead{$\log$ M$_{\rm gas,mol}$[M$_{\odot}$]  } 
& \colhead{$\log$ t$_{\rm dep}$[yr] } 
& \colhead{$\log$ M$_{\rm gas,mol}$[M$_{\odot}$] } &  \colhead{$\log$ t$_{\rm dep}$[yr] } & \colhead{$\log$ M$_{\rm gas,mol}$[M$_{\odot}$]  } & \colhead{$\log$ t$_{\rm dep}$[yr]  } \\
 & \multicolumn{2}{c}{\citet{Scoville2016}} & \multicolumn{2}{c}{\citet{Hughes2017}} &  \multicolumn{2}{c}{Gas-to-Dust Ratio}
 }
\startdata
DR3-195616 & $10.43^{+0.11}_{-0.11}$ & 
$8.63^{+0.11}_{-0.15}$ & $10.50^{+0.31}_{-0.31}$ &
$8.71^{+0.31}_{-0.31}$ & $10.34^{+0.22}_{-0.36}$ &
$8.54^{+0.22}_{-0.37}$ \\
DR1-209435 & $11.75^{+0.10}_{-0.13}$ & 
$9.16^{+0.12}_{-0.18}$ &  $11.60^{+0.32}_{-0.32}$ & 
$9.01^{+0.33}_{-0.33}$ & $11.52^{+0.22}_{-0.22}$ & 
$8.93^{+0.24}_{-0.24}$  \\
\enddata
\end{deluxetable*}

%% file: Table_Prop_in_figures.tex
\begin{deluxetable*}{lcccccccc}
\setlength{\tabcolsep}{7pt}
\tablecaption{Summary of stellar and gas properties. \label{tab:Figure_prop}}
\tablehead{
\colhead{UMG} & 
\colhead{$\log$ M$_\star$} & 
\colhead{SFR} & 
\colhead{sSFR} & 
\colhead{$\log$ M$_{\rm gas,mol}$} & 
\colhead{M$_{\rm gas,mol}$/M$_{\star}$ } &
\colhead{t$_{\rm dep}$} & 
\colhead{$\log$ M$_{\star,final}$} \\
\colhead{} & 
\colhead{[M$_\odot$/yr]} & 
\colhead{[M$_\odot$]} & 
\colhead{[Gyr$^{-1}$]} & 
\colhead{[M$_\odot$]} & 
 &
\colhead{[Gyr]} & 
\colhead{[M$_\odot$]}
}
\startdata
COS-DR3-195616 & 
$11.36^{+0.04}_{-0.05}$ & 
$62.47 \pm 9.59$ & 
$0.28 \pm 0.05$ & 
$10.43^{+0.11}_{-0.11}$ & 
$0.12^{+0.03}_{-0.03}$ &
$0.43 \pm 0.13$ & 
$11.38^{+0.04}_{-0.05}$ \\
COS-DR1-209435 & 
$11.38^{+0.10}_{-0.13}$ & 
$389.15 \pm 83.45$ & 
$1.62 \pm 0.53$ & 
$11.75^{+0.10}_{-0.13}$ & 
$2.33^{+0.83}_{-0.83}$ &
$1.44 \pm 0.48$ & 
$11.72^{+0.09}_{-0.12}$ \\
\enddata
\end{deluxetable*}

%% file: AGN_classification.tex
\begin{deluxetable*}{lcccccc}
\centering
\caption{AGN classification for two UMGs.
\label{tab: AGN_classification}}
\tablehead{
\colhead{UMG} & \colhead{X-ray AGN} & \colhead{Radio-AGN} & \colhead{SED f$_{\rm AGN} $} & \colhead{SED f$_{\rm AGN, MIR} $} & \colhead{AGN Bolometric Luminosity} & \colhead{Emission Line Ratios} \\
\colhead{} & \colhead{} & \colhead{} & \colhead{(RF 3--1000 $\mu$m)} & \colhead{(RF 1--30 $\mu$m)} & \colhead{ [erg s$^{-1}$] } & \colhead{} }
\startdata
DR3-195616 & Yes & Radio-quiet & $0.60\pm0.03$ &  0.73 & $4.1 \pm 0.4 \times 10^{45}$ & [OIII]5007/H$\beta = 6.22\pm1.16$ \\
DR1-209435 & No & Radio-quiet & $0.31\pm0.10$ &  0.89 &  $7.5 \pm 3.2 \times 10^{45}$ & [NII]6584/H$\alpha = 0.62\pm0.16$ \\
\enddata
\end{deluxetable*}

%% file: Table_CIGALE_input_t2.tex
\begin{deluxetable*}{lcll}
\tablecaption{Summary of \texttt{CIGALE} SED fitting model parameters. 
\label{tab:CIGALE_para}}
\tablehead{
\colhead{Component} & \colhead{Module} & \colhead{Parameter} & \colhead{Fitting Values} }
\decimalcolnumbers
\startdata
Star formation history & sfhdelayed & tau\_main (Gyr) & 100, 500, 1000, 3000, 5000 \\
{ } & { }  & age\_main (Myr) & 500, 1000, 1500, 2000, 2500\\
{ } & { } & age\_burst (Myr) & 10, 40, 70\\
{ } & { } & f\_burst & 0, 0.0001, 0.001, 0.01, 0.1, 0.3 \\
\hline
Simple stellar population & BC03 & IMF & Chabrier (2003) \\
{ } & { } & Metallicty & 0.02\\
\hline
Nebular Emission & { } & Gas metallicity & 0.005, 0.012, 0.02, 0.03\\ 
\hline
Dust attenuation & modified starburst & E\_BV\_lines &  0.1, 0,3, 0.5, 0.7, 1, 1.3, 1.5\\
{ } & { } & uv\_bump\_amplitude & 0.0, 1.5, 3\\
{ } & { } & powerlaw\_slope & -0.4, -0.2, 0\\
\hline
Dust emission & dl2014 & Mass fraction of PAH (qpah) & 0.47, 1.12, 1.77, 2.5, 4.58\\
{ } & { } & Minimum radiation field (umin) & 0.1, 1, 5, 10, 15\\
{ } & { } & Power-low index of radiation field intensity (alpha) & 1, 2, 3\\
\hline
AGN & skirtor2016 & viewing angle (i) & 30$^{\circ}$ (type 1), 70$^{\circ}$ (type 2)\\
{ } & { } & AGN contribution to IR luminosity (frac$_{\rm AGN}$) & 0.1 to 0.99 (in steps of 0.1)\\
{ } & { } & Polar dust color excess & 0.03, 0.1, 0.2, 0.3\\
\hline
Radio & {} & SF radio-IR correlation parameter (q$_{\rm IR}$) & 2.4, 2.5, 2.6, 2.7\\
{ } & { } & Radio-loudness parameter (R$_{\rm AGN}$) & 0.1 ,1, 5, 10, 50, 100
\enddata
\tablecomments{For parameters not listed here, we use the default values from \texttt{CIGALE v2022.1}.}
\end{deluxetable*}

%% file: Table_OH_vertical.tex
\begin{table}[htbp]
\centering
\caption{Oxygen abundance from \texttt{CIGALE} SED fitting, mass-metallicity relations, fundamental metallicity relation, and $N2$ index calibration. 
We take all MZRs/FMRs as reference calibrations to compute $\overline{\sigma_{\text{scale}}}$, defined as the average offset between the SED-derived metallicity and the values predicted by these relations. 
\label{tab: O_H_vertical}}
\begin{tabular}{lccc}
\hline
\hline
 Method  & Reference &  DR3-195616  & DR1-209435  \\
\hline 
SED & CIGALE & 8.614$^{+0.186}_{-0.332}$ & 8.881$^{+0.132}_{-0.191}$ \\
MZR & \citet{Genzel2015} & 8.626$^{+0.008}_{-0.010}$ & 8.667$^{+0.015}_{-0.022}$ \\
MZR & \citet{Maiolino2008} & 8.620$^{+0.011}_{-0.012}$ & 8.904$^{+0.016}_{-0.022}$ \\
MZR & \citet{Ma2016} & 8.654$^{+0.015}_{-0.017}$ & 8.753$^{+0.034}_{-0.044}$ \\
FMR & \citet{Mannucci2010} & -- & 8.753$^{+0.029}_{-0.031}$ \\
Calib. & PP04 N2 index & -- & 8.782$^{+0.180}_{-0.180}$ \\
\hline
Unc & $\overline{\sigma_{\text{scale}}}$ & 0.019 & 0.123 \\
\hline
\end{tabular}
\end{table}